\documentclass[conference]{IEEEtran}
\IEEEoverridecommandlockouts
\usepackage{cite}
\usepackage{amsmath,amssymb,amsfonts}
\usepackage{algorithmic}
\usepackage{graphicx}
\usepackage{textcomp}
\usepackage{subfigure}
\def\BibTeX{{\rm B\kern-.05em{\sc i\kern-.025em b}\kern-.08em
    T\kern-.1667em\lower.7ex\hbox{E}\kern-.125emX}}
\usepackage{url}
\usepackage{booktabs} 
\usepackage{colortbl}
\usepackage[ruled, vlined, linesnumbered]{algorithm2e}

\usepackage{multirow}
\usepackage{amsmath}
\usepackage{comment}
\usepackage{mathtools}
\usepackage{multirow}

\usepackage{colortbl}
\usepackage[dvipsnames]{xcolor}
\usepackage{balance}

\begin{document}

\title{Converting Your Thoughts to Texts: Enabling Brain Typing via Deep Feature Learning 
of 
EEG Signals}

\author{\IEEEauthorblockN{Xiang Zhang}
\IEEEauthorblockA{\textit{School of Computer Science \& Engineering} \\
\textit{University of New South Wales}\\
Sydney, Australia \\
xiang.zhang3@student.unsw.edu.au}
\and
\IEEEauthorblockN{Lina Yao}
\IEEEauthorblockA{\textit{School of Computer Science \& Engineering} \\
\textit{University of New South Wales}\\
Sydney, Australia \\
lina.yao@unsw.edu.au}
\and
\IEEEauthorblockN{Quan Z. Sheng}
\IEEEauthorblockA{\textit{Department of Computing} \\
\textit{Macquarie University}\\
Sydney, Australia  \\
michael.sheng@mq.edu.au}
\and
\IEEEauthorblockN{Salil S. Kanhere}
\IEEEauthorblockA{\textit{School of Computer Science \& Engineering} \\
\textit{University of New South Wales}\\
Sydney, Australia \\
salilk@cse.unsw.edu.au}
\and
\IEEEauthorblockN{Tao Gu}
\IEEEauthorblockA{\textit{School of Science} \\
\textit{RMIT University}\\
Melbourne, Australia \\
tao.gu@rmit.edu.au}
\and
\IEEEauthorblockN{Dalin Zhang}
\IEEEauthorblockA{\textit{School of Computer Science \& Engineering} \\
\textit{University of New South Wales}\\
Sydney, Australia \\
dalin.zhang@student.unsw.edu.au}
}

\maketitle

\begin{abstract}
An electroencephalography (EEG) based Brain Computer Interface (BCI) enables people to communicate with the outside world by interpreting the EEG signals of their brains to interact with devices such as wheelchairs and intelligent robots. More specifically, motor imagery EEG (MI-EEG), which reflects a subject's active intent, is attracting increasing attention for a variety of BCI applications.
Accurate classification of MI-EEG signals while essential for effective operation of BCI systems, is challenging due to the significant noise inherent in the signals and the lack of informative correlation between the signals and brain activities. 
In this paper, we propose a novel deep neural network based learning framework that affords perceptive insights into the relationship between the MI-EEG data and brain activities.
We design a joint convolutional recurrent neural network that simultaneously learns robust high-level feature presentations through low-dimensional dense embeddings from raw MI-EEG signals. We also employ an Autoencoder layer to eliminate various artifacts such as background activities. The proposed approach has been evaluated extensively on
a large-scale public MI-EEG dataset and a limited but easy-to-deploy dataset collected in our lab.
The results show that our approach outperforms a series of baselines and the 
competitive state-of-the-art methods, yielding a classification accuracy of 95.53\%.
The applicability of our proposed approach is further demonstrated with a practical BCI system for typing.
\end{abstract}

\begin{IEEEkeywords}
EEG, deep learning, brain typing, BCI
\end{IEEEkeywords}

\section{INTRODUCTION}
Brain-computer interface (BCI) systems have been widely studied for various real-world applications from mind-controlled service robots in the healthcare domain \cite{pinheiro2016wheelchair} to enriched video gaming in the entertainment industry \cite{Alomari2014}.  
As an important pathway between human brains and the outside world \cite{nguyen2015eeg}, BCI systems allow people to communicate or interact with external devices such as wheelchairs or service robots, through their brain signals.
Among the different types of brain signals, motor imagery Electroencephalography (MI-EEG) is especially popular and has demonstrated promising potential in discerning different brain activities in BCI systems. Motor imagery is a mental process where a subject {\em imagines} performing a certain action such as closing eyes or moving feet. Basically, EEG\footnote{In this paper, we will use the terms EEG and MI-EEG interchangeably.} 
is a method to analyze brain activities by measuring the voltage fluctuations of ionic current within the neurons of brains. In practice, electrodes are usually placed on the scalp for the measurement in a non-invasive and 
non-stationary way \cite{niedermeyer2005electroencephalography}. 

One of the most promising and widely discussed application of EEG-based BCI is to enable people to type via direct brain control \cite{akram2015efficient}.
In this paper, we aim at enabling a brain typing system by enhancing the decoding accuracy of EEG signals for wider range of brain activities (e.g., multi-class scenario). 
We envision a real-world implementation of such a system which can interpret the user's thoughts to infer typing commands in real-time.
 Motor disabled people would benefit greatly from such a system to express their thoughts and communicate with the outer world.

However, EEG signals fluctuate rapidly and are subject to various sources of noise including environmental noise such as lighting and electronic equipment.
Thus, the key issue concerning an EEG-based BCI system is to accurately interpret EEG signals so as to accurately understand the user's intent.
More specifically, 
the design of  a practical and effective BCI-system is faced with the following major challenges. First, EEG signals usually have very low signal-to-noise ratio \cite{repovs2010dealing}. 
As a result, EEG signals inherently lack sufficient spatial resolution and insight on 
activities of deep brain structures. 
Second, data pre-processing, parameter selection (e.g., filter type, pass band, segment window, and overlapping), 
and feature engineering (e.g., feature selection and extraction both in time domain and frequency domain) are all time-consuming and highly dependent on human expertise in the domain. 
Third, the state-of-the-art approaches can achieve an accuracy of at most 70$\sim$85\%, which though impressive is not sufficient for widespread adoption of this technology.
Fourth, existing research mainly focuses on discerning EEG signals under the binary classification situation and little work has been conducted on multi-class scenarios. Intuitively, the more scenarios an EEG-based control system can distinguish, 
the wider is its applicability in the real-world.

To tackle the aforementioned challenges, we propose a novel hybrid deep neural network that combines the benefits of both Convolutional Neural Networks (CNNs) \cite{sak2014long} and recurrent neural networks (RNNs) \cite{mikolov2010recurrent} for effective EEG signal decoding. Our model is capable of modeling high-level, robust and salient feature representations hidden in the raw EEG signal streams and capturing complex relationships within data via stacking multiple layers of information processing modules in a hierarchical architecture. Specifically, RNN is designed to model sequential information while CNN is well suited to extract higher-level spatial variations. In particular, a specific RNN architecture, named Long Short-Term Memory (LSTM), is designed to model temporal sequences and their long-range dependencies more accurately than conventional RNNs. In comparison, CNN is a typical feed-forward architecture and is able to extract higher-level features that are invariant to local spectral and temporal variations. 
The main contributions of this paper are highlighted as follows:
\begin{itemize}
\item We design a unified deep learning framework that leverages recurrent convolutional neural network to capture spatial dependencies of {\em raw} EEG signals based on features extracted by convolutional operations and temporal correlations through RNN architecture, respectively. Moreover, an Autoencoder layer is fused to cope with the possible incomplete and corrupted EEG signals to enhance the robustness of EEG classification.
\item We extensively evaluate our model using a public dataset and also a limited but easy-to-deploy dataset that we collected using an off-the-shelf EEG device. 
The experiment results illustrate that the proposed model achieves high-level of accuracy over both the public dataset (95.53\%) and the local dataset (94.27\%).
This demonstrates the consistent applicability of our proposed model. 
We have made our local dataset and the source code used in our evaluations available to the research community to encourage further research in this area and foster reproducibility of results.
\item We also present an operational prototype of a brain typing system based on our proposed model, which demonstrates the efficacy and practicality of our approach. A video demonstrating the system is made available \footnote{\url{http://1015xzhang.wixsite.com/mysite/demos}}. 
\end{itemize}

\begin{table}[!ht]
\centering
\begin{scriptsize}
\caption{The correlation coefficients matrix. Self, Cross, and PD separately denote self-similarity, cross-similarity and percentage difference.
}
\label{tab:c1}
\resizebox{\linewidth}{!}{\begin{tabular}{llllll|ccc}
\hline
\rowcolor[HTML]{C0C0C0} 
\textbf{Class}   & \textbf{0}      & \textbf{1}      & \textbf{2}      & \textbf{3}      & \textbf{4}      & \textbf{Self} & \textbf{Cross} & \textbf{PD} \\ \hline
0       & 0.4010 & 0.2855 & 0.4146 & 0.4787 & 0.3700 & 0.401           & 0.3872           & 3.44\%                \\
1       & 0.2855 & 0.5100 & 0.0689 & 0.0162 & 0.0546 & 0.51            & 0.1063           & 79.16\%               \\
2       & 0.4146 & 0.0689 & 0.4126 & 0.2632 & 0.3950 & 0.4126          & 0.2854           & 30.83\%               \\
3       & 0.4787 & 0.0162 & 0.2632 & 0.3062 & 0.2247 & 0.3062          & 0.2457           & 19.76\%               \\
4       & 0.3700 & 0.0546 & 0.3950 & 0.2247 & 0.3395 & 0.3395          & 0.3156           & 7.04\%                  \\ \hline
Range   & 0.1932 & 0.4938 & 0.3458 & 0.4625 & 0.3404 & 0.2038          & 0.2809           & 75.72\%               \\
Average & 0.3900 & 0.1870 & 0.3109 & 0.2578 & 0.2768 & 0.3939          & 0.2680           & 28.05\%               \\
STD     & 0.0631 & 0.1869 & 0.1334 & 0.1487 & 0.1255 & 0.0700          & 0.0932           & 27.33\%               \\ \hline
\end{tabular}
}
\end{scriptsize}
\vspace{-5mm}
\end{table}

\section{EEG CHARACTERISTICS ANALYSIS} 
\label{sec:eeg_characteristic_analysis}
The key point of the brain typing system is to precisely classify the user's intent signals. Although EEG signals have low signal-to-noise ratio and are sensitive to background brain activities and environmental factors, it is possible to recognize human intent by employing appropriate feature representation and classification.

To illustrate this point, we briefly analyze the similarities between EEG signals corresponding to  different intents and quantify them using \textit{Pearson correlation coefficient}. 
To be able to effectively interpret multiple classes of human intents, we assume that the EEG signals should meet the two hypotheses: 1) the intra-intent correlation coefficients should be consistently higher than inter-intent correlation coefficients;
2) the greater the difference between the intra-intent and inter-class correlation coefficients, the better classification results. 

According to the two hypotheses, we introduce two similarity concepts used in our measurement: the {\em self-similarity} and the {\em cross-similarity}. Self-similarity measures the similarity of EEG signals within the same intent. We randomly select several EEG data samples from the same intent and calculate the correlation coefficient of each possible \textit{pair of samples}. The self-similarity for the specific intent is measured as the average of all the sample pairs' correlation coefficients.
Cross-similarity
is defined to measure the similarity of two samples belonging to different EEG categories. For each specific intent, we 
measure the correlation coefficient of each possible \textit{intent pairs}. In this work, the EEG dataset (discussed in detail in Section~\ref{sub:setting}) contains 5 intents, hence there are a total of 20 intent pairs and 4 intent pairs of each specific intent. Finally, for each specific intent, the cross-similarity is the average of correlation coefficients of each intent pair. Also, we measure the correlation coefficients matrix for each specific subject and then calculate the average matrix (by calculating the mean value of all the matrix). For example, if there are 5 intents for a specific subject, we calculate a $5*5$ similarity matrix. In the matrix, $\rho_{\breve{i},\breve{j}}$ denotes the correlation coefficients between the sample of the intent $\breve{i}$ and the sample of the intent $\breve{j}$. 

Table~\ref{tab:c1} shows the correlation coefficients matrix and the corresponding statistical self- and cross-similarity. The last column (PD) denotes Percentage Difference between the self-similarity and cross-similarity. We can observe from the results that the self-similarity is always higher than the cross-similarity for all intents, which means that the intra-intent cohesion of the samples is stronger than the inter-intent cohesion. Moreover, the percentage difference has a noticeable fluctuation, which demonstrates that the intra-intent cohesion varies between different intents. The above analysis results 
justify the two hypotheses and lay the foundation for us to design appropriate feature representations and the classifier.

\begin{figure*}
\includegraphics[width=\linewidth]{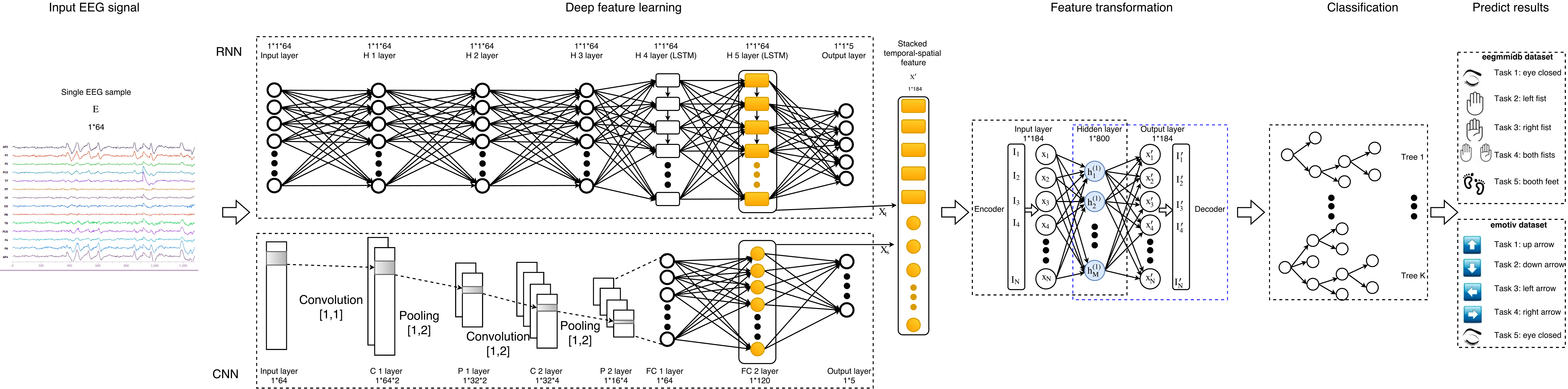}
\caption{The flow chart of the proposed approach. The input raw EEG data is a single sample vector denoted by $E_{\bar{i}} \in \mathbb{R}^K$ (take $K=64$ as an example). The \textit{C 1 layer} denotes the first convolutional layer, the \textit{C 2 layer} denotes the second convolutional layer, and so on. The same theory, the \textit{P 1 layer} denotes the first pooling layer; the \textit{FC 1 layer} denotes the first fully connected layer; the \textit{H 1 layer} denotes the first hidden layer. The \textit{stacked temporal-spatial feature} is generated by the \textit{FC 2 layer} in CNN and the \textit{H 5 layer} in RNN.}
\label{fig:overview}
\end{figure*}

\section{THE DEEP INTENT RECOGNITION MODEL}
\label{sec:approach}
This paper proposes a hybrid deep learning model to classify the raw EEG signal. In this section, we first provide an overview of the proposed approach and
 then present the technical details in subsequent sub-sections. 

\subsection{Overview}
\label{sub:overview}
Figure~\ref{fig:overview} illustrates the various steps involved.
The essential goal of our approach is to design a deep learning model that 
precisely classifies the user's intents based on EEG data.
 To obtain the useful and robust EEG features, we employ a parallel feature learning method which 
combines RNN
and 
CNN.
RNN is useful in extracting the internal memory to process arbitrary sequences of EEG signal while 
CNN is well-suited for the dimensional relevance representation. 
 In summary, we propose a hybrid approach which contains several components: the deep feature learning (Section~\ref{sub:rnn}), the feature transformation and the intent recognition (Section~\ref{sub:ae}). 


\subsection{Deep Feature Learning}
\label{sub:rnn}

We aim to learn the representations of the user's typing intent signal which is a 1-D vector (collected in one time-point). Let us represent the single input EEG signal as $E_{\bar{i}} \in \mathbb{R}^K$
 ($K=64$) 
 with $K$ is the number of dimensions in the EEG raw signal. Next, we feed $E_{\bar{i}}$ to the RNN structure and the CNN structure for temporal and spatial feature learning in parallel. At last, the learned temporal features $X_t$ and the spatial features $X_s$ are combined into the \textit{stacked feature} $X'$ for the latter feature transformation (Section~\ref{sub:ae}).

\subsubsection{RNN Feature Learning} 
\label{sec:temporal_feature_learning}
In the temporal feature processing part, the RNN structure is employed for its powerful ability for temporal feature extraction in time-series data.
RNN, which is one class of deep neutral network, is able to explore the feature dependencies over time through an internal state of the network, which allows it to exhibit dynamic temporal behavior. In this section, we take advantages of this trait to represent the temporal feature of the input EEG signal. 

We design an RNN model 
consisting of three components: one input layer, 5 hidden layers, and one output layer. 
There are two layers of Long Short-Term Memory (LSTM) \cite{zaremba2014recurrent} (shown as the rectangles in Figure~\ref{fig:overview}) cells among the hidden layers. 
Assume a batch of input EEG data contains $n_{bs}$ (generally called batch size) EEG samples and 
the total input data has the 3-D shape as $[n_{bs},1,64]$. Let the data in the $i$-th layer ($i=1,2, \cdots, 7$) be denoted by $X_i^r= \{X_{ijk}^r|j=1,2, \cdots, n_{bs}, k=1,2, \cdots, K_{i}\}, X_i^r \in \mathbb{R}^{[n_{bs},1,K_{i}]}$, where $j$ denotes the $j$-th EEG sample and $K_i$ denotes the number of dimensions in the 
$i$-th layer.

Assume that the weights between layer \textit{i} and layer \textit{i}+1 can be denoted by \(W_{i(i+1)}^r \in \mathbb{R}^{[K_i,K_{i+1}]}\), for instance, $W_{12}^r$ describes the weight between layer 1 and layer 2. $b_i^r \in \mathbb{R}^{K_i}$ denotes the biases of \textit{i}-th layer. 
The calculation between the $i$-th layer data and the $i+1$-th layer data can be denoted as
$$X_{i+1}^r=X_{i}^r*W_{i,i+1}^r+b_i^r$$

Please note the sizes of $X_{i}^r$, $W_{i,i+1}^r$ and $b_i^r$ must 
match. For example, in Figure~\ref{fig:overview}, the transformation between H1 layer and H2 layer, the sizes of $X_3^r$, $X_2^r$, $W_{[2,3]}$, and $b_2^r$ are separately $[1,1,64]$, $[1,1,64]$, $[64,64]$, and $[1,64]$.

The 5-th and 6-th layers in the designed structure are LSTM layers, so the calculation in these layers are implemented as follows:
\[f_i=sigmoid(T(X_{(i-1)j}^r,X_{(i)(j-1)}^r))\]
\[f_f=sigmoid(T(X_{(i-1)j}^r,X_{(i)(j-1)}^r))\]
\[f_o=sigmoid(T(X_{(i-1)j}^r,X_{(i)(j-1)}^r))\]
\[f_m=tanh(T(X_{(i-1)j}^r,X_{(i)(j-1)}^r))\]
\[c_{ij}=f_f\odot c_{i(j-1)}+f_i\odot f_m\]
\[X_{ij}^r=f_o\odot tanh(c_{ij})\]
where $f_i,f_f,f_o$ and $f_m$ represent the input gate, forget gate, output gate and input modulation gate 
separately, and $\odot$ denotes the element-wise multiplication. The $c_{ij}$ denotes the state (memory) in the $j$-th LSTM cell in 
the $i$-th layer, which is the most significant part to explore the time-series relevance between samples. 
The $T(X_{(i-1)j}^r,X_{(i)(j-1)}^r)$ denotes the operation as follows:
$$X_{(i-1)j}^r*W+X_{(i)(j-1)}^r*W'+b$$
where $W$, $W'$ and $b$ denote the corresponding weights and biases.

At last, we obtain the RNN 
prediction results $X_7^r$ and employ cross-entropy as the cost function. The cost is optimized by the AdamOptimizer algorithm \cite{kingma2014adam}. 
$X_6^r$ is the data in the second last layer, which has a directly linear relationship with the output layer and the prediction results. 
 If the predicted results have high accuracy, $X_6^r$ is enabled to directly map to the sample label space and has the better representative of the input EEG sample. Therefore, we regard $X_6^r$ as the temporal feature extracted by the RNN structure and call it $X_t$.

\subsubsection{CNN Feature Learning} 
\label{sub:cnn}
While RNN is good in exploring the temporal (inter-sample) relevance, it is unable to appropriately decode spatial feature (intra-sample) representations. To exploit the spatial connections between different features in each specific EEG signal, we design a CNN structure. The CNN structure is comprised of three categories of components: the convolutional layer, the pooling layer, and the fully connected layer. The convolutional layer contains a set of \textit{filters} to convolve with the EEG data and then 
through the feature pooling and non-linear transformation to extract the geographical features. 
CNN is well-suited to extract the spatial relevance of the 2-D input data efficiently. In this paper, 
we implement the CNN on the 1-D EEG data. 
As shown in Figure~\ref{fig:overview}, the designed CNN is stacked in the following order: the input layer, the first convolutional layer, the first pooling layer, the second convolutional layer, the second pooling layer, the first fully connected layer, the second fully connected layer, and the output layer.

The input is the same EEG data as the RNN. 
The input EEG single sample $E_{\bar{i}}$ has shape $[1,64]$. 
Suppose the data in the $i$-th layer ($i=1,2, \cdots, 8$) is denoted by $X_i^c,X_i^c \in \mathbb{R}^{[1,K_{i}^c,d_i]}$, where $K_i^c$ and $d_i$ separately denote the dimension number and the depth in the $i$-th layer. The data in the first layer only has depth 1 and $X_1^c= E$. We choose the convolutional filter with size $[1,1]$ and the stride size $[1,1]$ in the first convolution. The stride denotes the x-movements and y-movements distance of the filter. The padding method is selected as same shape zero-padding, which results in the sample shape keeping constant in the convolution calculation. The depth of EEG sample transfers to 2 through the first convolutional layer, so the shape of $X_2^c$ is $[1,64,2]$. 

The pooling layer is a non-linear down-sampling transformation layer. There are several pooling options, with \textit{max pooling} being the most popular \cite{nagi2011max}. The max pooling layer scans through the inputs along with a sliding window with a designed stride.
Then it outputs the maximum value in every sub-region that the window is scanned. The pooling layer reduces the spatial size of the input EEG features and also 
prevents overfitting. In the first pooling layer (the 3-th layer in CNN), we choose the $[1,2]$ window and $[1,2]$ stride. 
The maximum in each $[1,2]$ window will be output to the next layer. The pooling does not change the depth and the shape of $X_3^c$ is $[1,32,2]$. Similarly, the second convolutional layer chooses $[1,2]$ filter and $[1,1]$ stride and gets the shape as $[1,32,4]$. The second pooling layer selects $[1,2]$ window and $[1,2]$ stride and obtains the shape as $[1,16,4]$.

In the full connected layer, the high-level reasoning features, extracted through previous convolutional and pooling layers, are unfolded to a flattened vector. For example, the data of the second pooling layer ($X_5^c$ with shape $[1,16,4]$) is flattened to the vector with shape $[1,64]$ ($X_6^c$). Then the output data can be calculated by following the regular neural network operation:
$$X_7^c=T(X_6^c)$$
$$X_8^c=softmax(T(X_8^c))$$

At last, we 
have the CNN results $X_8^c$ and employ the cross-entropy as the cost function. The cost is optimized by the AdamOptimizer algorithm. 
$X_7^c$ has a directly linear relationship with the output layer and the predicted results. Therefore, we regard 
$X_7^c$ as the spatial feature extracted by the CNN structure and call it $X_s$.

In summary, the temporal features $X_t$ and the spatial features $X_s$ are learned through the parallel RNN and CNN structures. Both of them have the 
direct linear relationship with the EEG sample label, which means that they
 represent the temporal and spatial features of the input EEG sample if both RNN and CNN 
have 
high classification accuracy. Next, we combine the two feature vectors into a flattened stacked vector,
$X'= \{X_t : X_s\}$.

\subsection{Feature Adaptation} 
\label{sub:ae}

Next, we design a 
feature adaptation 
method to map the stacked features to a correlative new feature space which can fuse the temporal and spatial features together and highlight the useful information.

To do so, we introduce the Autoencoder layer \cite{nguyen2015eeg} to further interpret EEG signals, which is an unsupervised approach to learning effective features. 
The Autoencoder is trained to learn a compressed and distributed representations for the stacked EEG feature $X'$. The input of Autoencoder is the stacked temporal and spatial feature $X'$. Assume $h$, $\acute{X'}$ denote the hidden layer and output layer data, respectively.


The data transformation procedure 
is described as the 
following:
$$h=W_{en}X'+b_{en}$$

$$\acute{X'}=W_{de}h+b_{de}$$
where $W_{en}$, $W_{de}$, $b_{en}$, $b_{de}$ 
denote the weights and biases in the encoder and the decoder. 

The cost function measures the difference between $X'$ and $\acute{X'}$ as MSE (mean squared error) which is back-propagated to 
the algorithm to adjust the weights and biases. The error is optimized by the RMSPropOptimizer \cite{hinton2012rmsprop}.
The data in the hidden layer $h$ is the transferred feature, which is output to the classifier. Finally, the Extreme Gradient Boosting) classifier (XGBoost) is employed  \cite{chen2016xgboost} to classify the EEG streams. It fuses a set of classification and regression trees (CART) and 
exploits detailed information from the input data. It builds multiple trees and each tree has its leaves and corresponding scores. 
\section{EXPERIMENTS}
\label{sec:experiment}

In this section, we evaluate the proposed deep learning model using a public dataset and a local dataset collected by ourselves. 
At first, a public EEG dataset (called \textit{eegmmidb}) is used to assess our proposed deep learning model. The experimental settings (Section~\ref{sub:setting}), the overall comparison with the state-of-the-art methods (Section~\ref{sub:overall_comparison}), the parameter tuning (Section~\ref{sec:parameter_tuning}), and the efficiency analysis (Section~\ref{sub:experiments_and_results}) are separately reported in this section. In addition, we 
evaluate our model on a local dataset for demonstrating the good adaptability of proposed method (the collected EEG dataset is called \textit{emotiv}) and present the corresponding results
 (Section~\ref{sub:case_study}).


\subsection{Experiment Setting}
\label{sub:setting}

We select the widely used EEG data from PhysioNet eegmmidb (\textit{EEG
motor movement/imagery database}) database\footnote{\url{https://www.physionet.org/pn4/eegmmidb/}}.
This data is collected using the BCI200 EEG system\footnote{\url{http://www.schalklab.org/research/bci2000}} \cite{schalk2004bci2000} which records the brain signals using 64 channels at a sampling rate of 160Hz. The subject is asked to wear the EEG device and sit in front of a computer screen and perform certain typing actions in response to hints that appear on the screen. The researchers have carefully annotated the EEG data to correspond to the actions undertaken by the subject, which are available from the PhysioBank ATM\footnote{\url{https://www.physionet.org/cgi-bin/atm/ATM}}. For our experiments, we select a total of 280,00 labeled EEG samples collected from 10 subjects (28,000 samples per subject). Each sample is a vector of 64 elements, each of which corresponds to one channel of the EEG data. The subjects performed 5 actions which are labeled as 0 to 4, as shown in Table~\ref{tab:table2}.

\begin{table}[!tb]
\begin{scriptsize}
\centering
\caption{The motor imagery tasks and labels
 and the corresponding typing command in the brain typing system}
\label{tab:table2}
\resizebox{\linewidth}{!}{\begin{tabular}{lllllll}
\hline
\rowcolor[HTML]{C0C0C0}
\textbf{Dataset} & \textbf{Item} & \textbf{Task 1} & \textbf{Task 2} & \textbf{Task 3} & \textbf{Task 4} & \textbf{Task 5} \\ \hline
\multirow{2}{*}{eegmmidb} & intent & eye closed & left hand & right hand & both hands & both feet \\
 & label & 0 & 1 & 2 & 3 & 4 \\ 
\multirow{2}{*}{emotiv} & intent & up arrow & down arrow & left arrow & right arrow & eye closed \\
 & label & 0 & 1 & 2 & 3 & 4 \\ \hline
 & command & up & down & left & right & confirmation \\ \hline
\end{tabular}
}
\vspace{-4mm}
\end{scriptsize}
\end{table}

To evaluate the performance of the classified results, 
we use several typical evaluation 
metrics such as accuracy, precision, recall, F1 score, ROC (Receiver Operating Characteristic) curve, and AUC (Area Under the Curve). 

\subsection{Overall Comparison}
\label{sub:overall_comparison}
In this section, we report the performance 
study and then demonstrate the efficiency of our approach by comparing with the state-of-the-art methods and other
independent deep learning algorithms. 
Recall that the proposed approach is a hybrid model which uses RNN and CNN for feature learning, the AE layer for feature transformation, and the XGBoost classifier for intent recognition. In this experiment, the EEG data is randomly divided into two parts: the training dataset (21,000 samples) and the testing dataset (7,000 samples). The accuracy of our method is calculated as the average of 5 runs on 10 subjects. 

\begin{figure*}[ht]
\centering
\begin{minipage}[b]{0.3\linewidth}
\centering
\includegraphics[width=\textwidth]{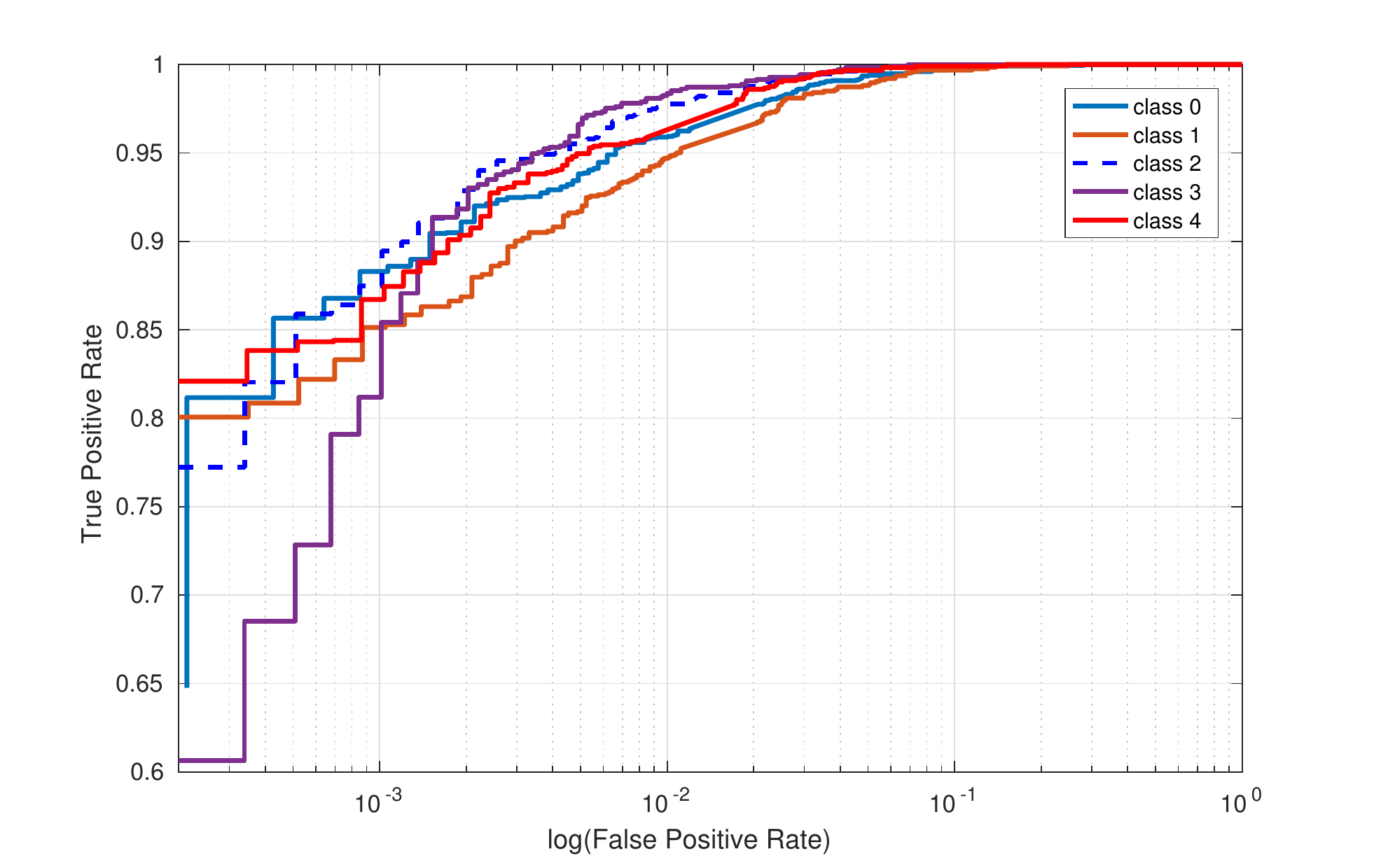}
\caption{The ROC curves of the 5-class classification. Note that X-axis is the logarithmic of the False Positive Rate. 
}
\label{fig:roc}
\end{minipage}
\hspace{4mm}
\begin{minipage}[b]{0.3\linewidth}
\centering
\includegraphics[width=\textwidth]{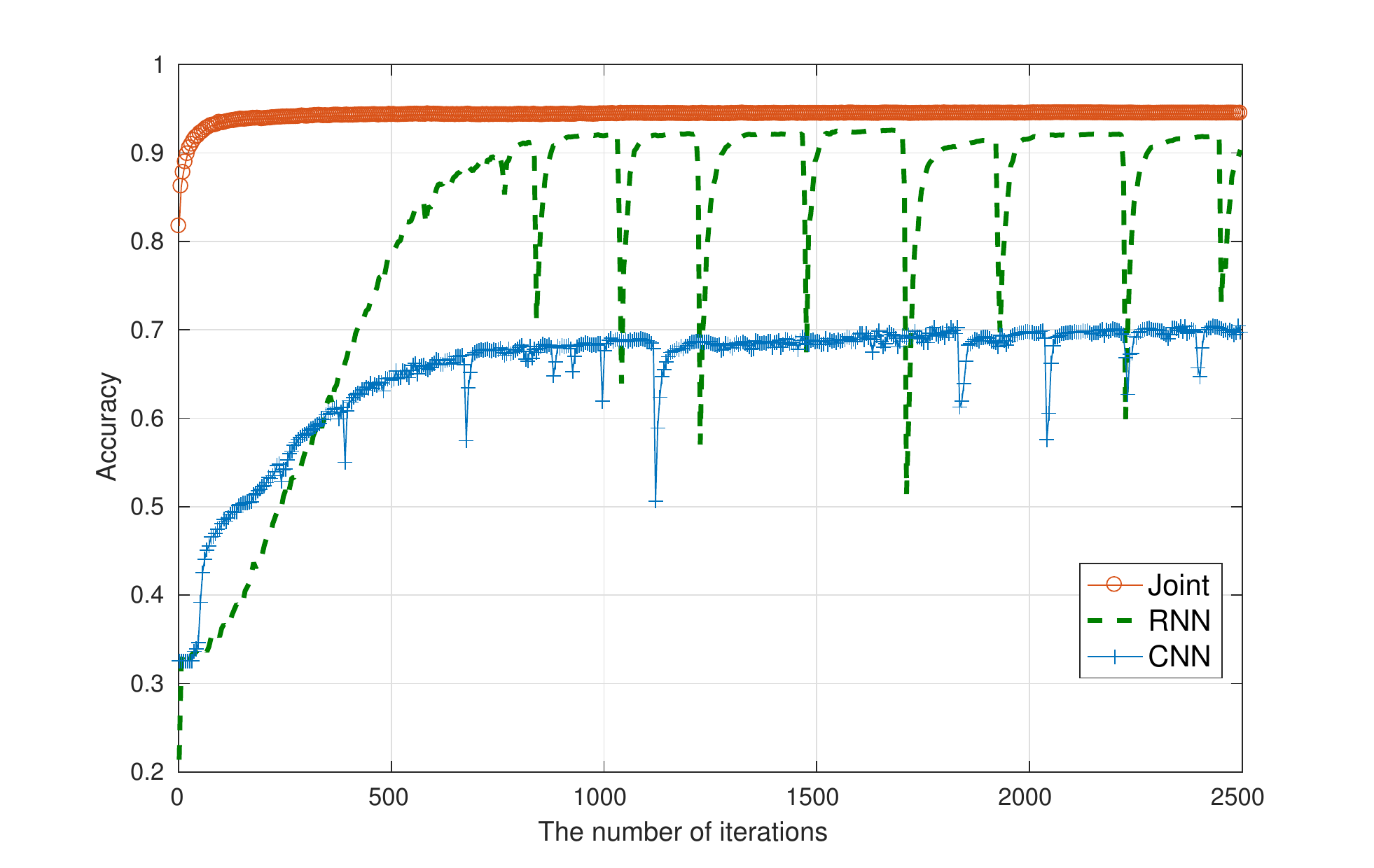}
\caption{The relationship between the testing accuracy and the number of iterations. 
}
\label{fig:acc}
\end{minipage}
\hspace{2mm}
\begin{minipage}[b]{0.34\linewidth}
\centering
\includegraphics[width=\textwidth]{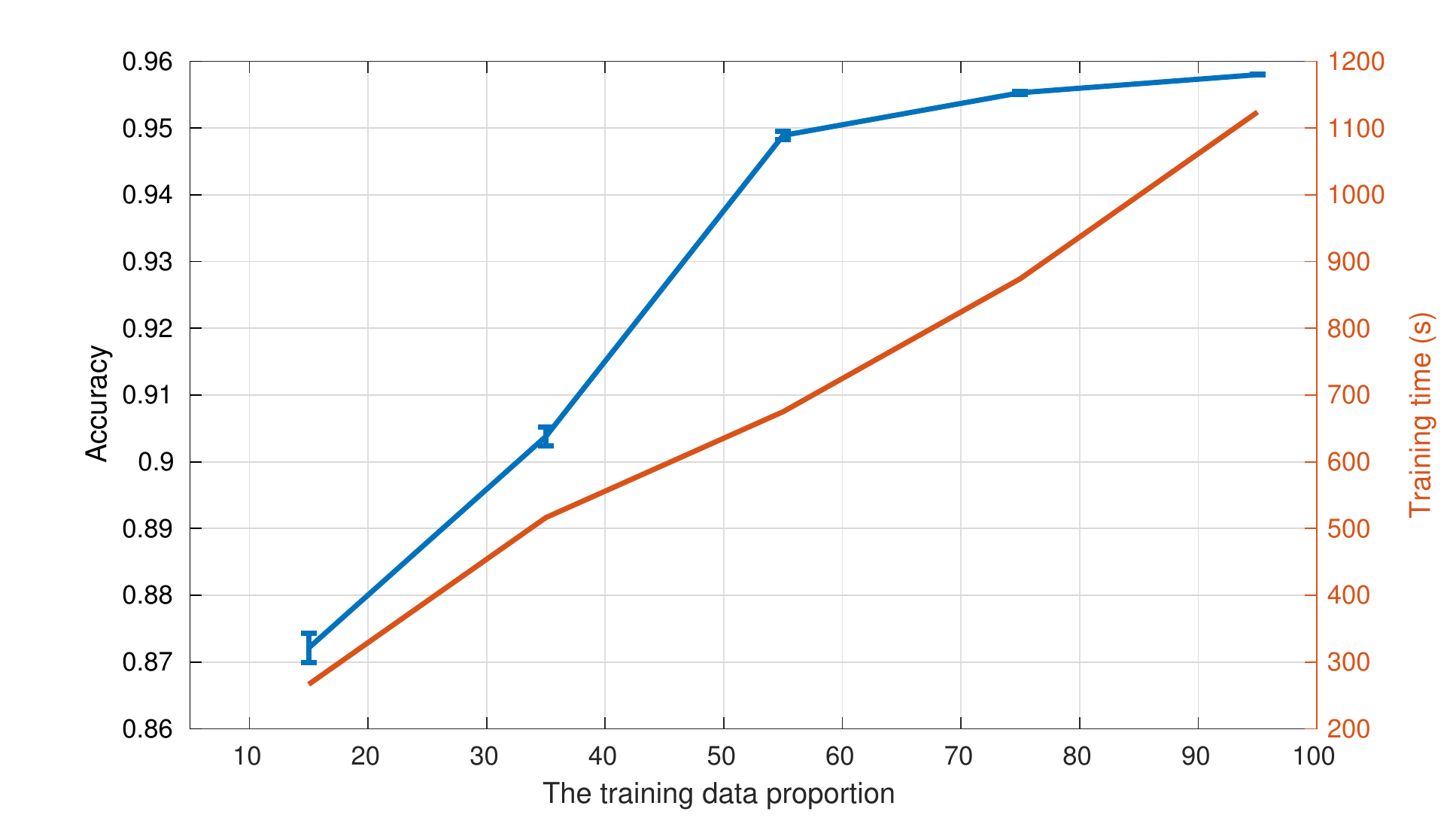}
\caption{The relationship between the classification accuracy and the training data proportion. 
}
\label{fig:datasize}
\end{minipage}
\vspace{-4mm}
\end{figure*}

Firstly, we report that our approach achieves the classification accuracy of \textbf{0.9553}. 
To take a closer look at the result, the detailed confusion matrix and classification reports are presented in Table~\ref{tab:confusion_matrix}. 
We can observe that for every class, our approach achieves an an average precision no lower than 0.939. 
Figure~\ref{fig:roc} shows the ROC curves of the 5 classes. 

Additionally, the accuracy comparison between our method and other state-of-the-art and baselines are listed in Table~\ref{tab:comparison}. Wavelet transform \cite{Alomari2014,sun2016classification,alomari2014eeg,tolic2013classification,or2016classification} and independent component analysis (ICA) \cite{major2017effects,sita2013feature} are state-of-the-art methods to process EEG signals. The Deep Neural Network \cite{major2017effects,shenoy2015shrinkage,tolic2013classification} and Linear discriminant analysis \cite{sita2013feature} are applied to classify the EEG data.
In addition, the key parameters of the baselines are listed here: KNN (k=3), Linear SVM ($C=1$), RF ($n=500$), LDA ($tol=10^{-4}$), and AdaBoost ($n=500, lr=0.3$).
  The results show that our method achieves the significantly higher accuracy of \textbf{0.9553} than all the state-of-the-art methods. Our method also 
performs better than other deep learning methods such as RNN or CNN. Moreover, compared with the 
most existing EEG classification research which focuses on binary classification, our method works in multi-class scenario but still achieves a high-level of accuracy.


\begin{table}[]
\begin{scriptsize}
\centering
\caption{The confusion matrix of 5-class classification}
\label{tab:confusion_matrix}
\resizebox{\linewidth}{!}{\begin{tabular}{lllllll|llll}
\hline
\rowcolor[HTML]{C0C0C0} 
 &  & \multicolumn{5}{l|}{\cellcolor[HTML]{C0C0C0}\textbf{Ground Truth}} & \multicolumn{4}{l}{\cellcolor[HTML]{C0C0C0}\textbf{Evaluation}} \\ \hline
 &  & 0 & 1 & 2 & 3 & 4 & Precision & Recall & F1 & AUC \\
 & 0 & \cellcolor[HTML]{FE0000}2062 & 19 & 23 & 18 & 22 & 0.9618 & 0.9380 & 0.9497 & 0.9982 \\
 & 1 & 17 & \cellcolor[HTML]{34FF34}1120 & 19 & 15 & 20 & 0.9404 & 0.9084 & 0.9241 & 0.9977 \\
 & 2 & 13 & 13 & \cellcolor[HTML]{F8FF00}1146 & 14 & 11 & 0.9574 & 0.9257 & 0.9413 & 0.9990 \\
 & 3 & 10 & 5 & 7 & \cellcolor[HTML]{96FFFB}1162 & 10 & 0.9732 & 0.9028 & 0.9367 & 0.9990 \\
\multirow{-6}{*}{\begin{tabular}[c]{@{}l@{}}Predict\\ Lable\end{tabular}} & 4 & 18 & 21 & 15 & 23 &\cellcolor[HTML]{329A9D} 1197 & 0.9396 & 0.9392 & 0.9394 & 0.9987 \\
Total &  & 2120 & 1178 & 1210 & 1232 & 1260 & 4.7723 & 4.6140 & 4.6911 & 4.9926 \\
Average &  &  &  &  &  &  & \textbf{0.9545} & \textbf{0.9228} & \textbf{0.9382} & \textbf{0.9985} \\ \hline
\end{tabular}
}
\end{scriptsize}
\end{table}

\begin{table}[]
\centering
\begin{scriptsize}
\caption{Performance comparison with the state of the art methods. 
RF denotes Random Forest and 
LDA denotes Linear Discriminant Analysis.}
\label{tab:comparison}
\begin{tabular}{lllcl}
\hline
\rowcolor[HTML]{C0C0C0} 
 & \textbf{Index} & \textbf{Methods} & \textbf{Binary/Multi} & \textbf{Acc}  \\ \hline
\begin{tabular}[c]{@{}l@{}}State\\ of the art\end{tabular} & 1 & Almoari \cite{Alomari2014} & \multirow{5}{*}{Binary} & 0.7497 \\
 & 2 & Sun \cite{sun2016classification} &  & 0.65 \\
 & 3 & Mohammad \cite{alomari2014eeg} &  & 0.845 \\
 & 4 & Major \cite{major2017effects} &  & 0.68 \\
 & 5 & Shenoy \cite{shenoy2015shrinkage} &  & 0.8206 \\
 & 6 & Tolic \cite{tolic2013classification} &  & 0.6821 \\
 & 7 & Rashid \cite{or2016classification} &  & 0.92 \\
 & 8 & Ward \cite{Ward2016} & Multi (3) & 0.8 \\
 & 9 & Sita \cite{sita2013feature} & Multi (3) & 0.8724 \\
 & 10 & Pinheiro \cite{pinheiro2016wheelchair} & Multi (4) & 0.8505 \\ \hline
\multirow{7}{*}{Baselines} & 8 & KNN & \multirow{8}{*}{Multi (5)} & 0.8769 \\
 & 11 & SVM &  & 0.5082 \\
 & 12 & RF &  & 0.7739 \\
 & 13 & LDA &  & 0.5127 \\
 & 14 & AdaBoost &  & 0.3431 \\
 & 15 & RNN &  & 0.9325 \\
 & 16 & CNN &  & 0.8409 \\
 & 17 & Ours &  & \textbf{0.9553} \\ \hline
\end{tabular}
\vspace{-3mm}
\end{scriptsize}
\end{table}
\begin{table*}[]
\centering
\begin{scriptsize}
\caption{The recognition accuracy of 10 subjects under different feature learning methods. The improvement represents the increase amplitude of our method over the maximum of RNN and CNN feature learning methods.
 }
\label{tab:extraction}
\begin{tabular}{llllllllllllll}
\hline
\rowcolor[HTML]{C0C0C0} 
\textbf{Feature learning} & \textbf{S1} & \textbf{S2} & \textbf{S3} & \textbf{S4} & \textbf{S5} & \textbf{S6} & \textbf{S7} & \textbf{S8} & \textbf{S9} & \textbf{S10} & \textbf{Range} & \textbf{average} & \textbf{std}    \\ \hline
RNN & 0.9005 & 0.8928 & 0.9506 & 0.9264 & 0.9487 & 0.9427 & 0.9098 & 0.9293 & 0.9643 & 0.8498 & 0.1145 & 0.9215 & 0.0341 \\
CNN & 0.9021 & 0.5938 & 0.9395 & 0.9659 & 0.9013 & 0.9942 & 0.9273 & 0.6177 & 0.9310 & 0.6358 & 0.4004 & 0.8409 & 0.1580 \\
RNN+CNN & 0.9390 & 0.9186 & 0.9784 & 0.9736 & 0.9967 & 0.9832 & 0.9675 & 0.9245 & 0.9758 & 0.8954 & 0.1013 & \textbf{0.9553} & 0.0335 \\
Improvement & 0.0369 & 0.0258 & 0.0278 & 0.0077 & 0.0480 & -0.0110 & 0.0402 & -0.0048 & 0.0115 & 0.0456 & 0.0590 & 0.0228 & 0.0209 \\ \hline
\end{tabular}
\end{scriptsize}
\end{table*}
To demonstrate the advantage of our proposed hybrid model for better learning of robust features from raw EEG data,  we also compare our method (joint RNN and CNN) with the independent deep feature learning method (RNN, CNN). All extracted features are classified by a XGBoost classifier.
The experimental results 
are listed in Table~\ref{tab:extraction}, where we 
can see that our approach outperforms RNN and CNN in classification accuracy by 3.38\% and 11.44\%, respectively.
Our approach also achieves the lowest standard deviation and range, implying that it is
more stable and reliable. Note that the RNN on its own (RNN works as both feature extract method and classifier) without feature representation achieves a higher accuracy of 0.9325 (in Table~\ref{tab:comparison}) than the \textit{RNN+AE+XGBoost} method (RNN works as feature extract method), which exhibits an accuracy of 0.9215.
 This shows that the RNN represented features are unsuitable for other classifiers and the inappropriate use of AE may \textit{decrease} the signal quality.  Figure~\ref{fig:acc} illustrates separately the accuracy changes along with the training iterations under three categories of feature learning methods. Three curves (in Figure~\ref{fig:acc}) show that the proposed joint method converges to its high accuracy in fewer iterations than independent RNN and CNN.
The learned features are fed into the AE for further processing and finally classified by the XGBoost classifier.

\subsection{Parameter Tuning} 
\label{sec:parameter_tuning}
In this section, we conduct a series of empirical studies for analyzing the impact of various parameters on the classification accuracy of the proposed approach. We extensively explore the impact of the following key factors: the training data size, the RNN learning rate, the CNN learning rate, the AE learning rate, the XGBoost learning rate, the AE hidden neuron size, and the classifier. 
We next investigate the impact of varying the data used for training on the accuracy of our model. The results are illustrated in Figure~\ref{fig:datasize}. As expected, the accuracy increases as more data is available for training. Our method achieves an accuracy of 95\% when 55\% of the available data set is used for training. There is a only a marginal improvement in accuracy with the inclusion of additional training data. Also observe that we can achieve an accuracy of 87\% with only 15\% of training data. This indicates that our approach is less dependent on the training data size. 
The time required for training the model is shown on the right vertical axis in Figure~\ref{fig:datasize} and as expected varies linearly with the size of the training data.


\begin{figure}
\centering
\subfigure[RNN learning rate]{
  \label{fig:rnn_lr}
  \includegraphics[width=0.44\linewidth]{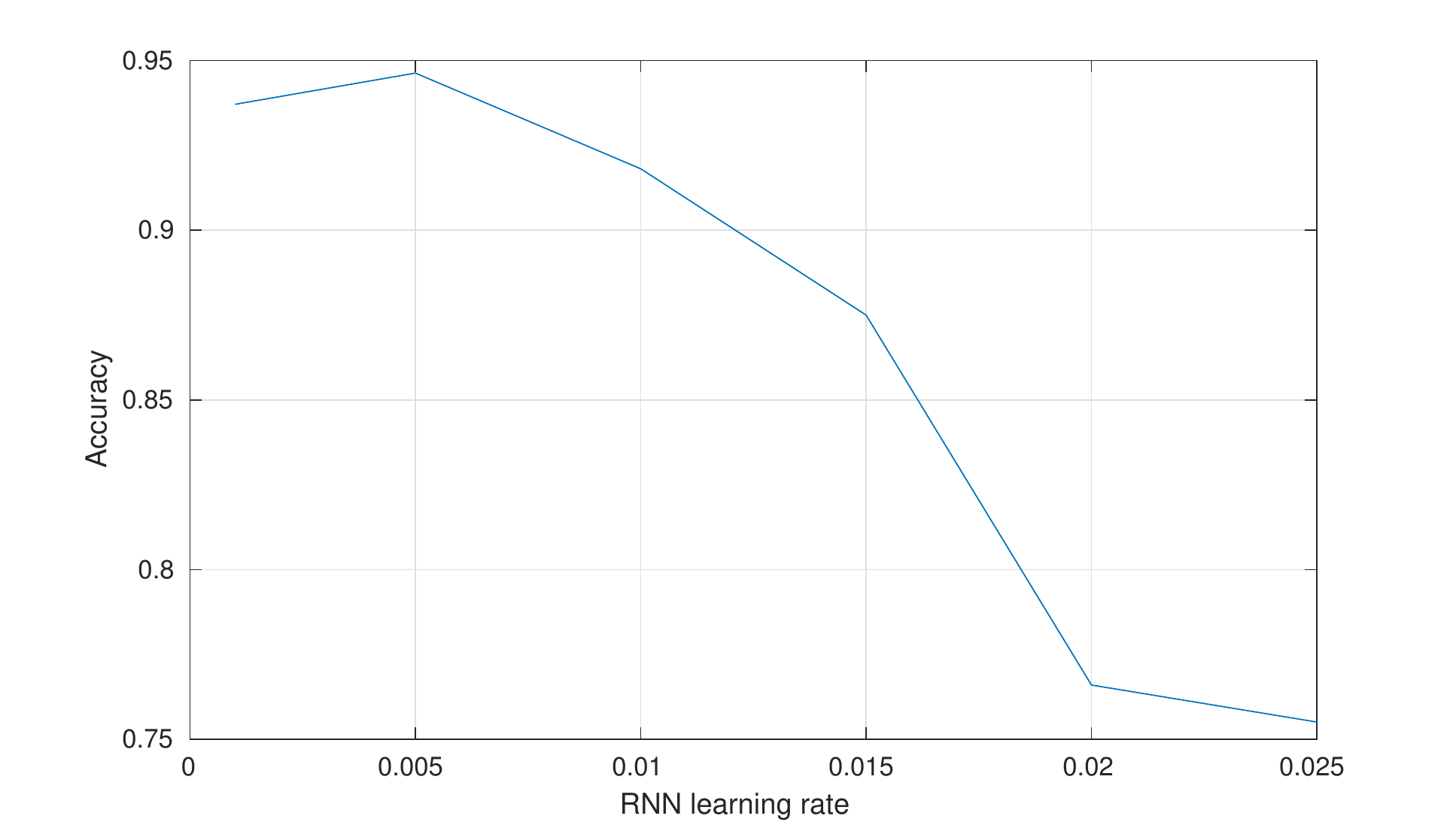}
}
\subfigure[CNN learning rate]{
  \label{fig:cnn_lr}
  \includegraphics[width=0.44\linewidth]{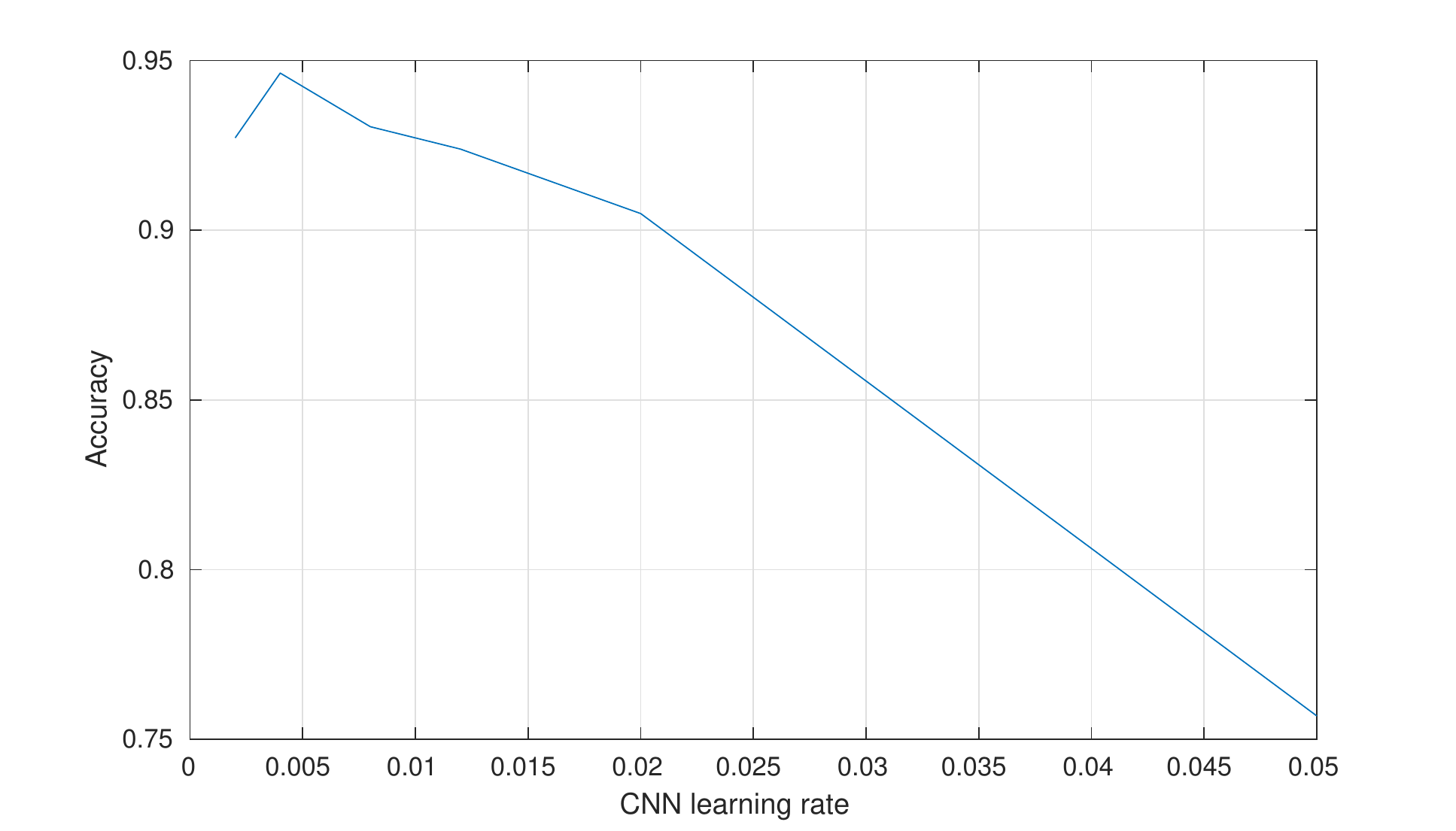} 
}
\subfigure[AE learning rate]{
  \label{fig:ae_lr}
  \includegraphics[width=0.44\linewidth]{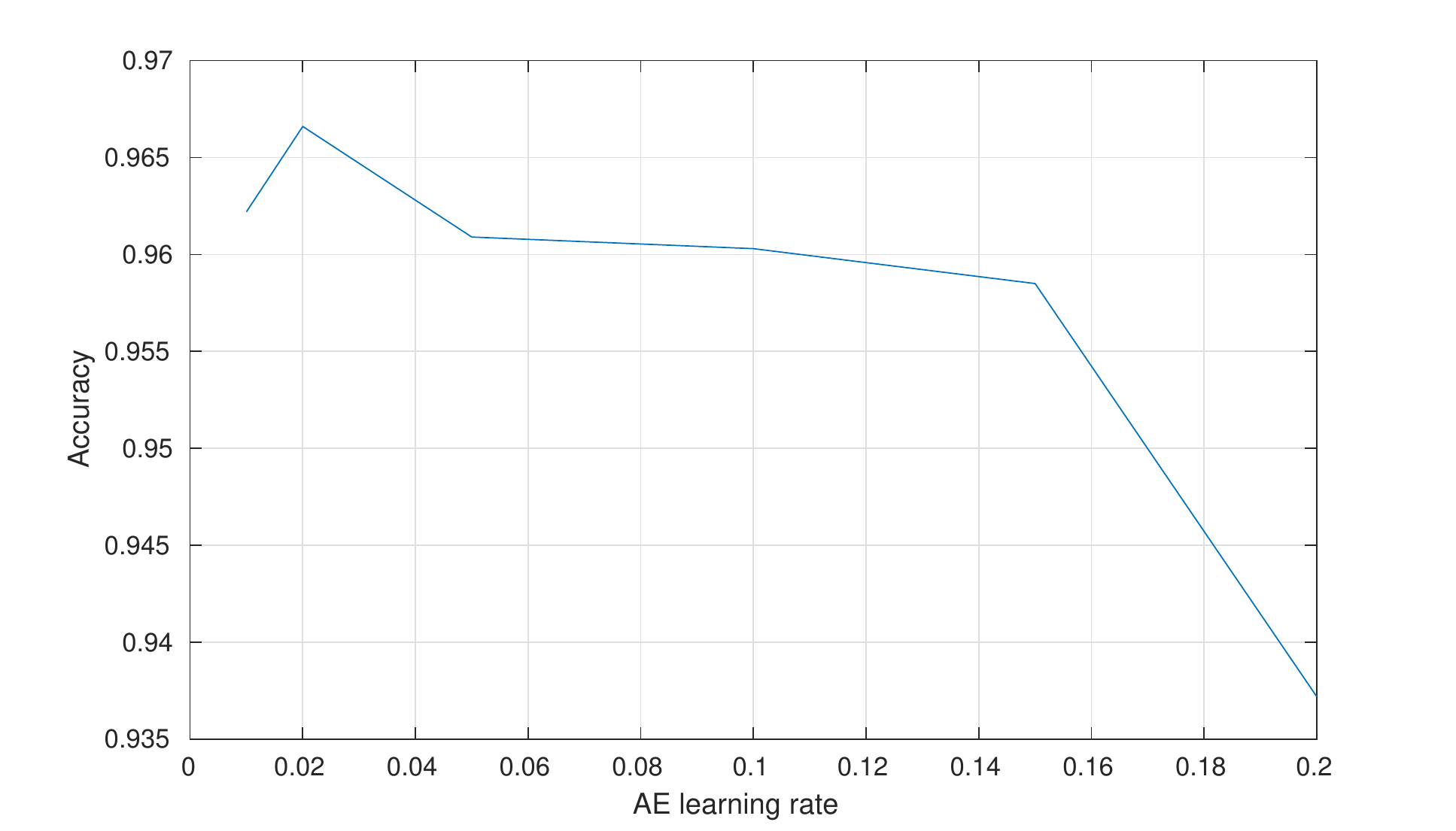} 
}
\subfigure[XGBoost learning rate]{
  \label{fig:xgb_lr}
  \includegraphics[width=0.44\linewidth]{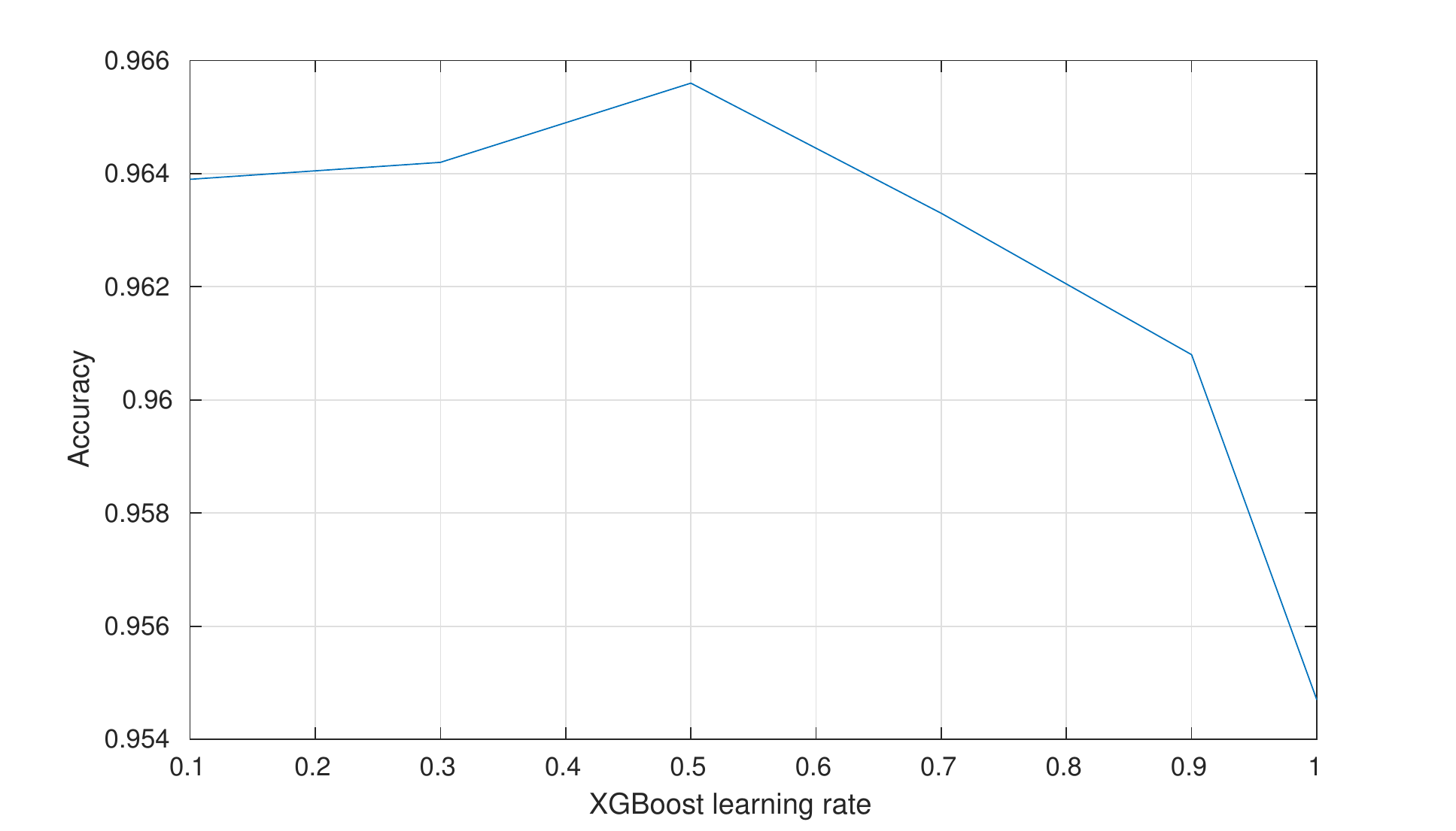} 
}
\subfigure[AE hidden neuron size]{
  \label{fig:ae_hidden}
  \includegraphics[width=0.44\linewidth]{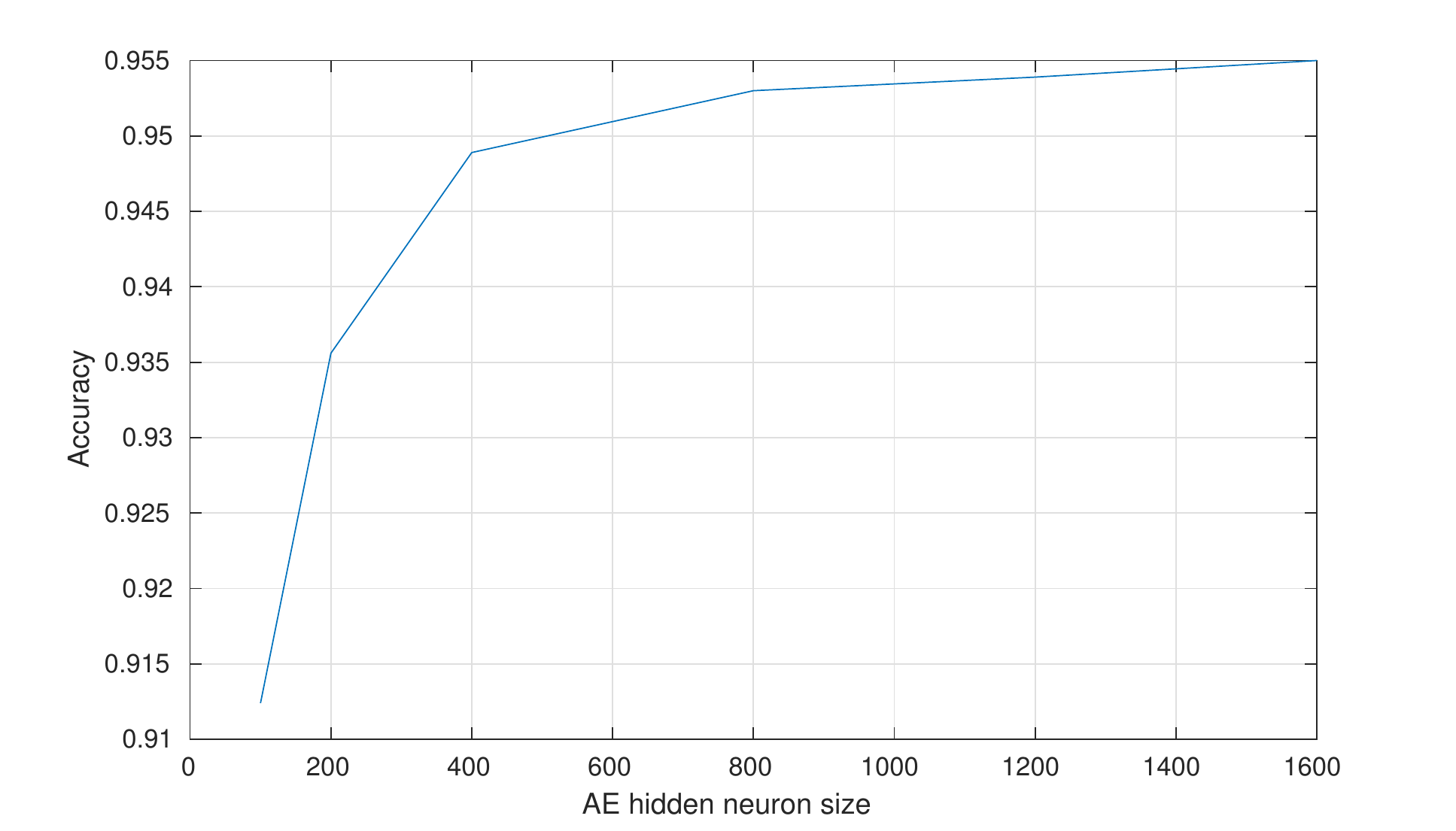} 
}
\subfigure[Classifier]{
  \label{fig:classifier}
  \includegraphics[width=0.44\linewidth]{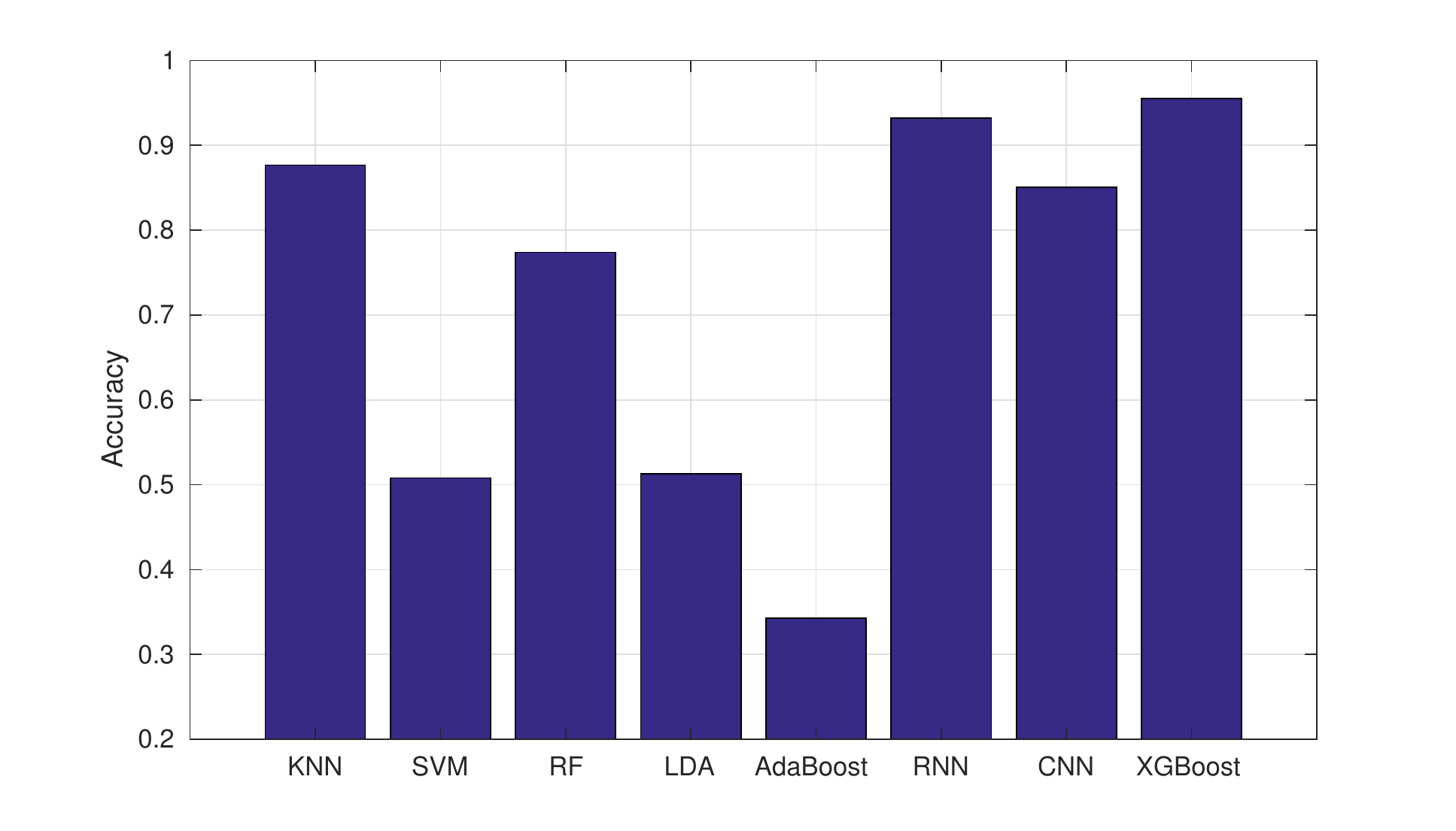} 
}
\caption{The classification accuracy with different hyper-parameter settings}
\label{fig:tuning}
\vspace{-5mm}
\end{figure}

Figure~\ref{fig:rnn_lr} to Figure~\ref{fig:xgb_lr} show that the proposed approach performs differently under different learning rates in each component.
 We choose the appropriate learning rates as 0.005, 0.004, 0.002, and 0.5 for RNN, CNN, AE, and XGBoost, respectively. 
Figure~\ref{fig:ae_hidden} illustrates that the more hidden neurons in AE, the better classification results. Therefore, we choose 800 neurons as a trade-off between the accuracy and efficiency. 
Figure~\ref{fig:classifier} shows that the XGBoost classifier outperforms other classifiers and achieves the highest classification accuracy over the same features refined by \textit{RNN+CNN+AE}.
It should be noted that 
all the not mentioned hyper-parameters are set as default value except those shown in Table~\ref{tab:parameter}. 

\begin{table}[!tb]
\centering
\begin{scriptsize}
\caption{Hyper-Parameter Setting. For instance, RNN contains one input layer (64 neurons), 5 hidden layers (64 neurons each layer), and one output layer (5 neurons).}
\label{tab:parameter}
\begin{tabular}{lll}
\hline
\rowcolor[HTML]{C0C0C0} 
\multicolumn{2}{l}{\cellcolor[HTML]{C0C0C0}\textbf{Hyper-parameter}} & \textbf{Value} \\ \hline
 & Layer & 7=1+5+1 \\
 & Neuron size & 64*1+64*5+5*1 \\
 & Iterations & 2500 \\
 & Batch size & 7000 \\
 & Learning rate & 0.005 \\
 & Activation function & Soft-max \\
 & Cost function & Cross entropy \\
\multirow{-8}{*}{RNN} & Regularization & $\ell_2$ norm ($\lambda=0.004$) \\ \hline
 & Layer & 8 \\
 & Input neuron size & 64 \\
 & 1st convolutional & Filter [1,1],stride [1,1], depth 2 \\
 & 1st pooling &  Window [1,2], stride [1,2] \\
 & 2nd convolutional & Filter [1,2],stride [1,1], depth 4 \\
 & 2nd pooling & Window [1,2], stride [1,2] \\
 & Padding method & Zero-padding \\
 & Pooling methods & Max \\
 & Activation function & ReLU \\
 & 1st fully connected & 64 \\
 & 2nd fully connected & 120 \\
 & Output neuron size & 5 \\
 & Iterations & 2500 \\
 & Batch size & 7000 \\
 & Learning rate & 0.004 \\
 & Activation function & Softmax \\
 & Cost function & Cross entropy \\
\multirow{-18}{*}{CNN} & Regularization & $\ell_2$ norm ($\lambda=0.001$) \\ \hline
 & Layer & 1+1+1 \\
 & Neuron size & 184+800+184 \\
 & Iterations & 400 \\
 & Learning rate & 0.01 \\
\multirow{-5}{*}{AE} & Cost function & MSE \\ \hline
 & Objective & Multi:softmax \\
 & Learning rate & 0.5 \\
 & max\_depth & 6 \\
\multirow{-4}{*}{Classifier} & Iterations & 500 \\ \hline
\end{tabular}
\end{scriptsize}
\vspace{-5mm}
\end{table}

\subsection{Efficiency Analysis} 
\label{sub:experiments_and_results}
\begin{figure}
\centering
\subfigure[Training time]{
  \label{fig:train_time}
  \includegraphics[width=0.44\linewidth]{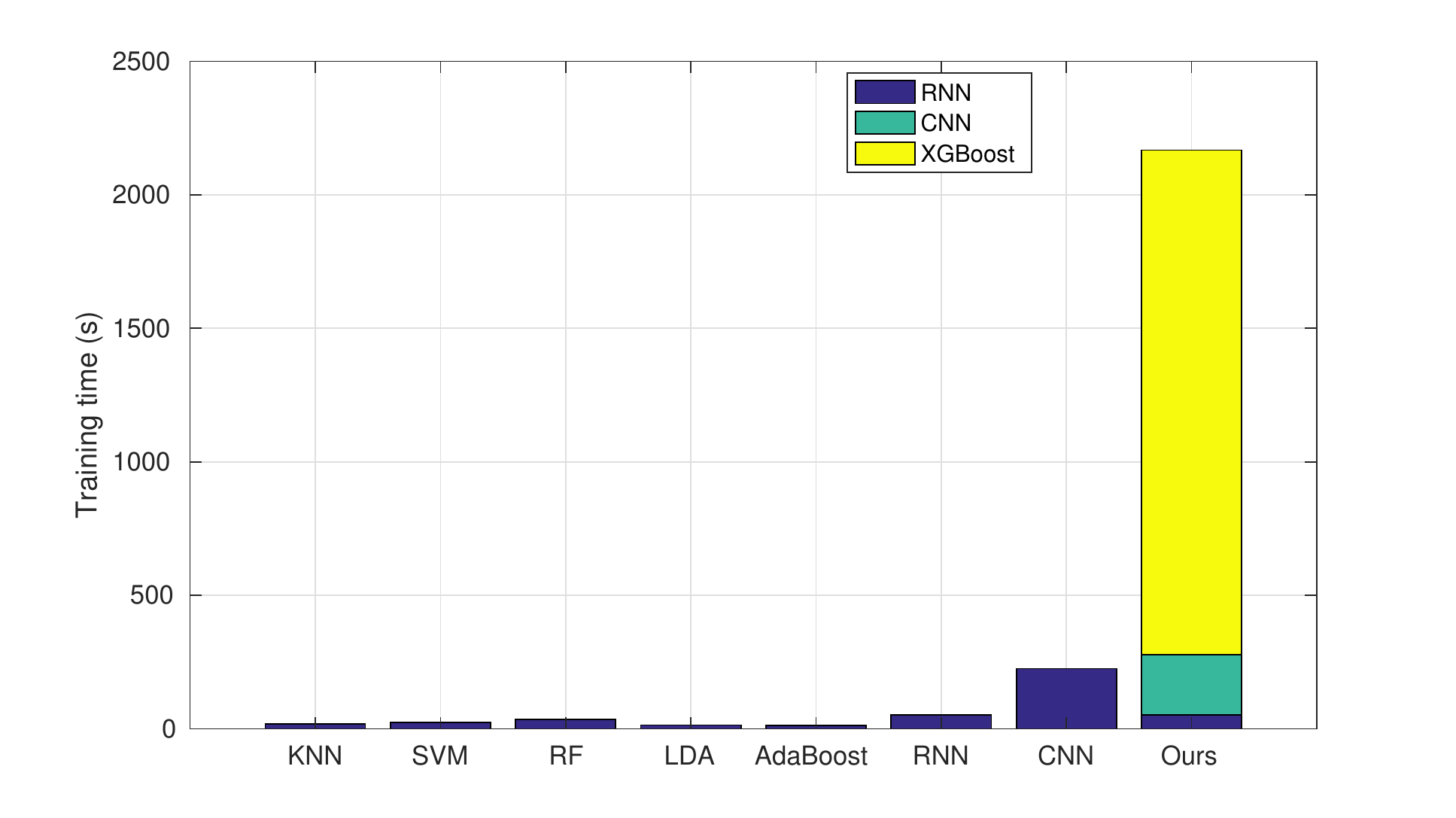}
}
\subfigure[Testing time]{
  \label{fig:test_time}
  \includegraphics[width=0.44\linewidth]{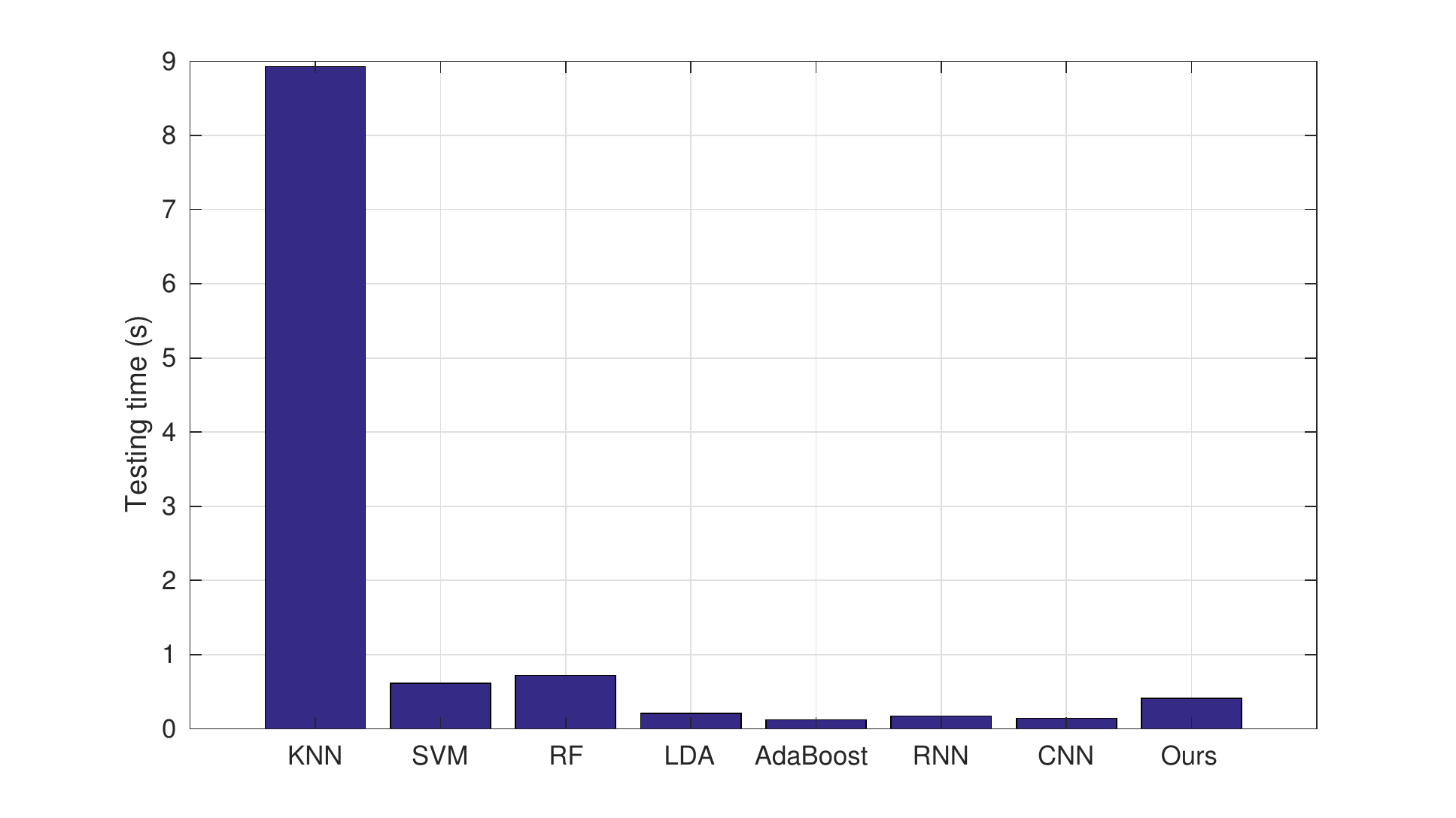} 
}
\caption{The training time and testing time comparison}
\label{fig:times}
\end{figure}

Generally, deep learning algorithms require substantial time to execute. This can limit their suitability for BCI applications (e.g., typing) which typically require close to real-time performance. 
For instance, the practical deployment of a BCI system 
could be limited by its recognition time-delay if it takes two minutes to recognize the user's intent. 
In this section, we will focus on the running time of our approach and compare it to the widely used baselines. The results are shown in Figure~\ref{fig:times}.

We first illustrate the time required to train the model in Figure~\ref{fig:train_time}. Our model requires 2,000 seconds for training, which is significantly longer than other baseline approaches. A breakdown of the training time required for the 3 components, namely, RNN, CNN and XGBoost is also shown. XGBoost requires the most training time as the result of its gradient boosting structure. However, training is a one-time operation. For practical considerations, the execution time of an algorithm during testing is what matters most.
Figure~\ref{fig:test_time} shows that the testing time of our approach is less than 1 second, which is similar with other baselines 
(except KNN which requires 9 seconds). In summary, the proposed approach takes very short testing time
 although it requires 
more time to train the model. Reducing the training time of our approach will be part of our future work.

\subsection{Adaptability Evaluation on Local EEG Dataset}
\label{sub:case_study}
\begin{figure}
\centering
\subfigure[EEG collection]{
  \label{fig:eeg_c}
  \includegraphics[width=0.4\linewidth]{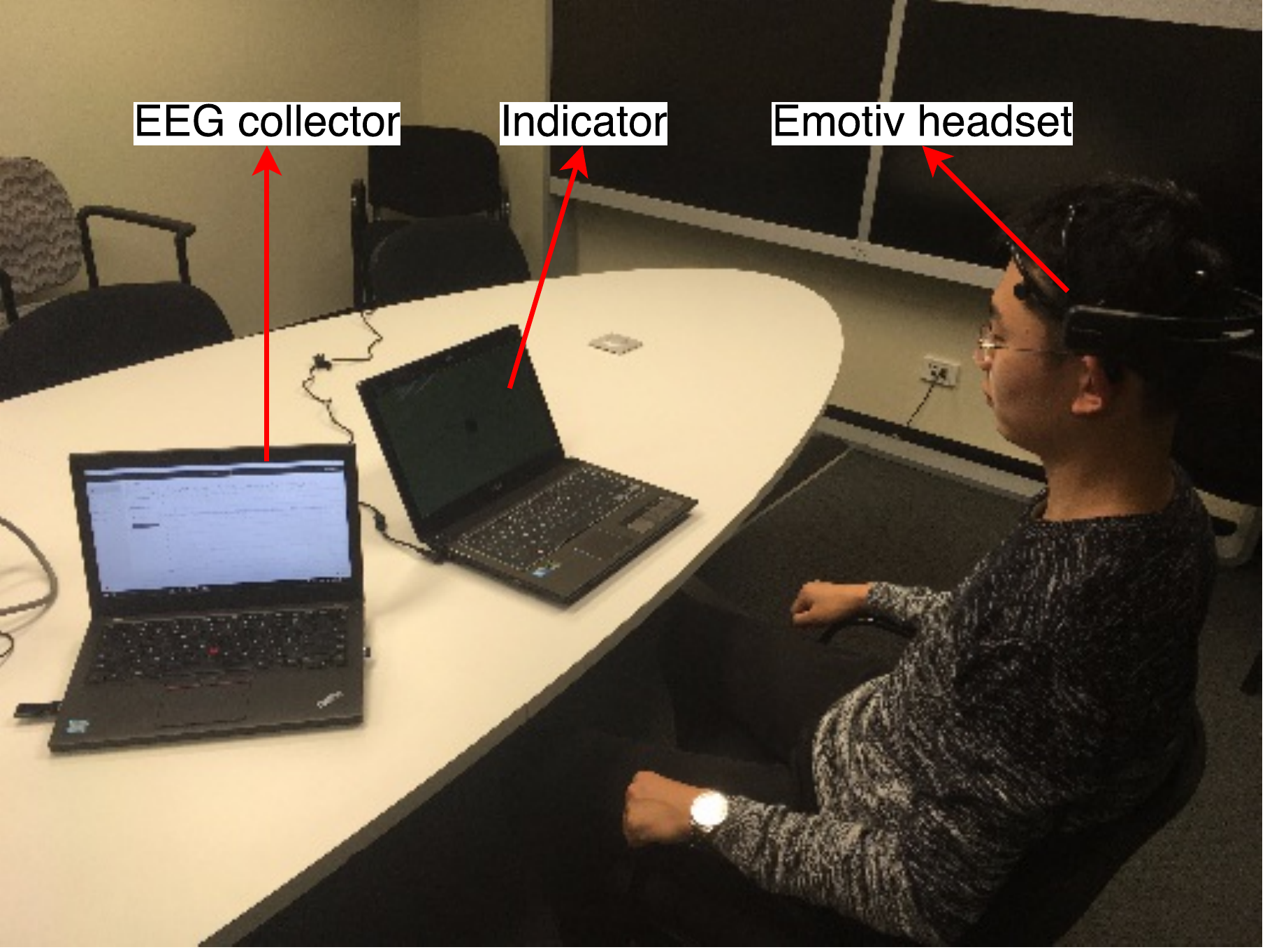}
}
\subfigure[EEG raw data]{
  \label{fig:eeg_r}
  \includegraphics[width=0.5\linewidth]{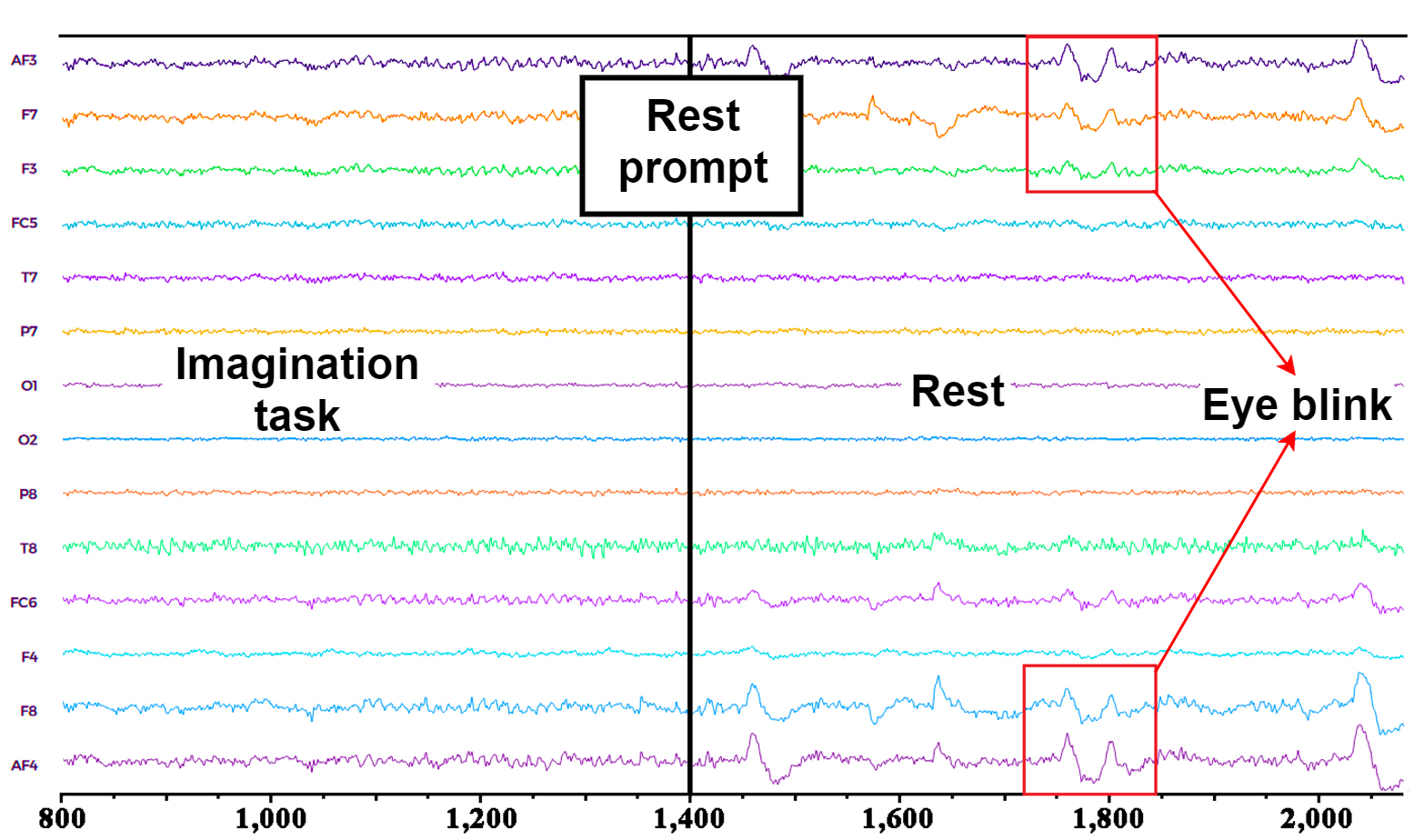} 
}
\caption{EEG collection and the raw data. The emotiv dataset only consists of the \textit{imagination task} data since the \textit{rest} state data is contaminated by eye blink and other noises.}
\label{fig:EEG_collection}
\end{figure}
To examine the adaptability and consistency of our model, we further evaluate our proposed model on a limited but easy-to-deploy dataset. We conduct the EEG collection by using a portable and easy-to-use commercialized EEG headset, Emotiv Epoc+ headset. The headset contains 14 channels and the sampling rate is 128 Hz. The local dataset can be accessed from this link\footnote{\url{https://drive.google.com/open?id=0B9MuJb6Xx2PIM0otakxuVHpkWkk}}. Compared to the BCI 2000 system (64 channels) used for construct the eegmmidb dataset, our local equipment (Emotiv headset) only contains 14 channels and is much easier to be deployed in a natural environment. 


\subsubsection{Experimental Setting} 
\label{sec:section_name}
This experiment is carried on by 7 subjects (4 males and 3 females) aged from 23 to 26. During the experiment, the subject wearing the \textit{Emotiv Epoc+}\footnote{\url{https://www.emotiv.com/product/emotiv-epoc-14-channel-mobile-eeg/}} EEG collection headset, faces the computer screen and focuses on the corresponding \textit{hint} which appears on the screen (shown in Figure~\ref{fig:EEG_collection}). The 
brain activities and labels used in this paper are listed in Table~\ref{tab:table2}. In summary, this experiment contains 241,920 samples with 34,560 samples for each subject. In order to distinguish with the aforementioned \textit{eegmmidb} dataset, 
we name this dataset as \textit{emotiv}.

\begin{table}[!tb]
\centering
\caption{The confusion Matrix and the evaluation over \textit{emotiv} dataset}
\label{tab:case_confusion}
\resizebox{0.5\textwidth}{!}{
\begin{tabular}{llllllllllll}
\hline
\rowcolor[HTML]{C0C0C0}
\textbf{} & \multicolumn{6}{l}{\textbf{Ground truth}} & \multicolumn{4}{l}{\textbf{Evaluation}} \\ \hline
 &  & 0 & 1 & 2 & 3 & 4 & Precision & Recall & F1 & AUC \\ 
 & 0 & \cellcolor[HTML]{F8A102}1608 & 35 & 51 & 21 & 18 & 0.9279 & 0.9415 & 0.9346 & 0.9982 \\
 & 1 & 18 & \cellcolor[HTML]{32CB00}1602 & 21 & 29 & 15 & 0.9507 & 0.9357 & 0.9432 & 0.9977 \\
 & 2 & 29 & 31 & \cellcolor[HTML]{FE0000}1642 & 27 & 22 & 0.9377 & 0.9437 & 0.9407 & 0.9990 \\
 & 3 & 19 & 25 & 11 & \cellcolor[HTML]{38FFF8}1615 & 36 & 0.9467 & 0.9428 & 0.9447 & 0.9990 \\
\multirow{-6}{*}{\begin{tabular}[c]{@{}l@{}}Predicted \\ Label\end{tabular}} & 4 & 34 & 19 & 15 & 21 & \cellcolor[HTML]{32CB00}1676 & 0.9496 & 0.9485 & 0.9490 & 0.9987 \\
Total &  & 1708 & 1712 & 1740 & 1713 & 1767 & 4.7126 & 4.7122 & 4.7123 & 4.9926 \\
Average &  &  &  &  &  &  & 0.9425 & 0.9424 & 0.9425 & 0.9985 \\ \hline
\end{tabular}
}
\end{table}

\subsubsection{Recognition Results and Comparison} 
\label{sec:recognition_results}
For each participant, the training set contains 25,920 samples and the testing set contains 8,640 samples. 
The experiment parameters are the same as listed in Table~\ref{tab:parameter}. The proposed approach achieves the 5-class classification accuracy of \textbf{0.9427}. The confusion matrix and evaluation is reported in Table~\ref{tab:case_confusion}. 

\begin{figure}[!tb]
\centering
\subfigure[Accuracy]{
  \label{fig:t2}
  \includegraphics[width=0.45\linewidth]{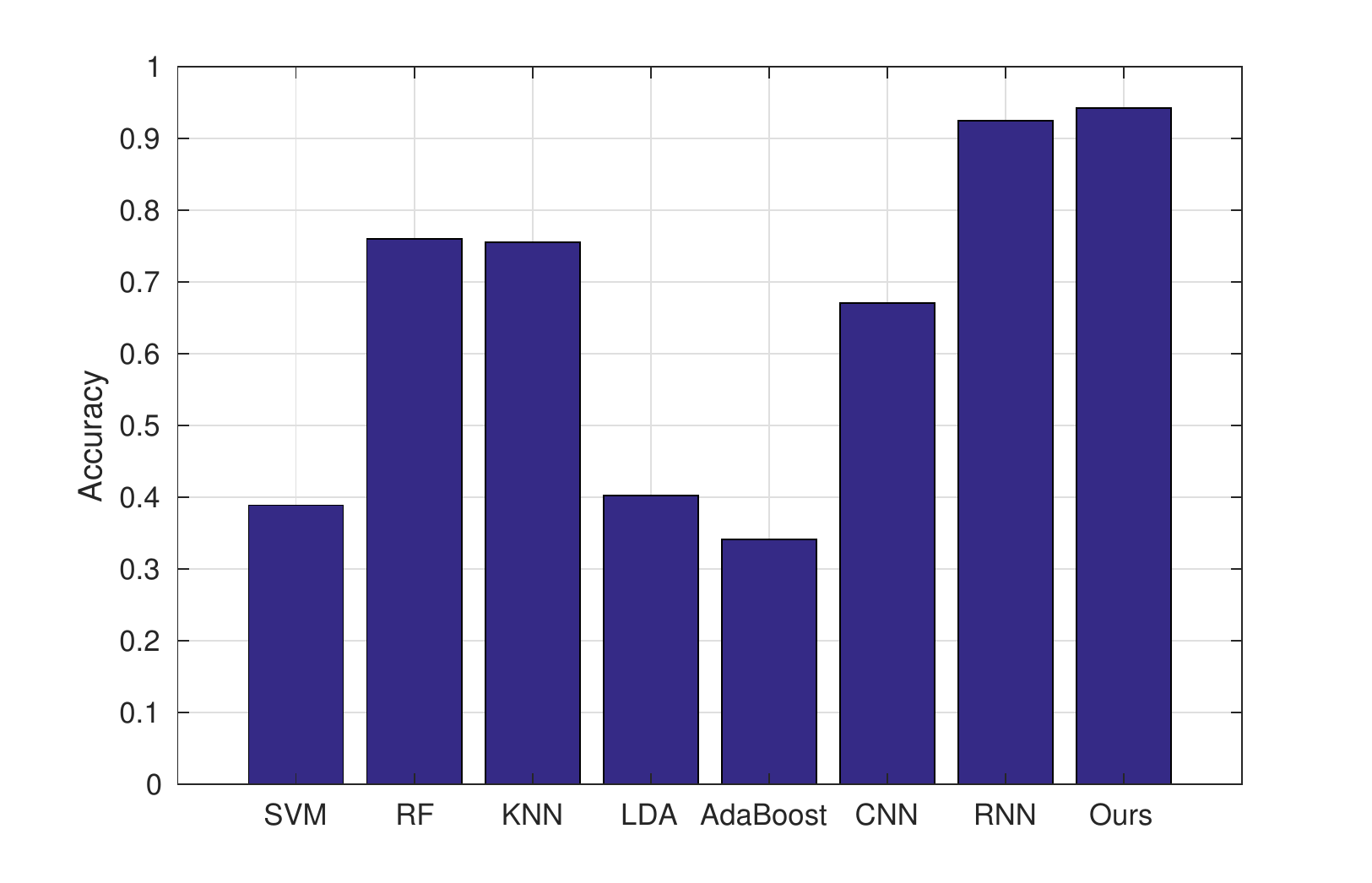}
}
\subfigure[Testing time]{
  \label{fig:t}
  \includegraphics[width=0.45\linewidth]{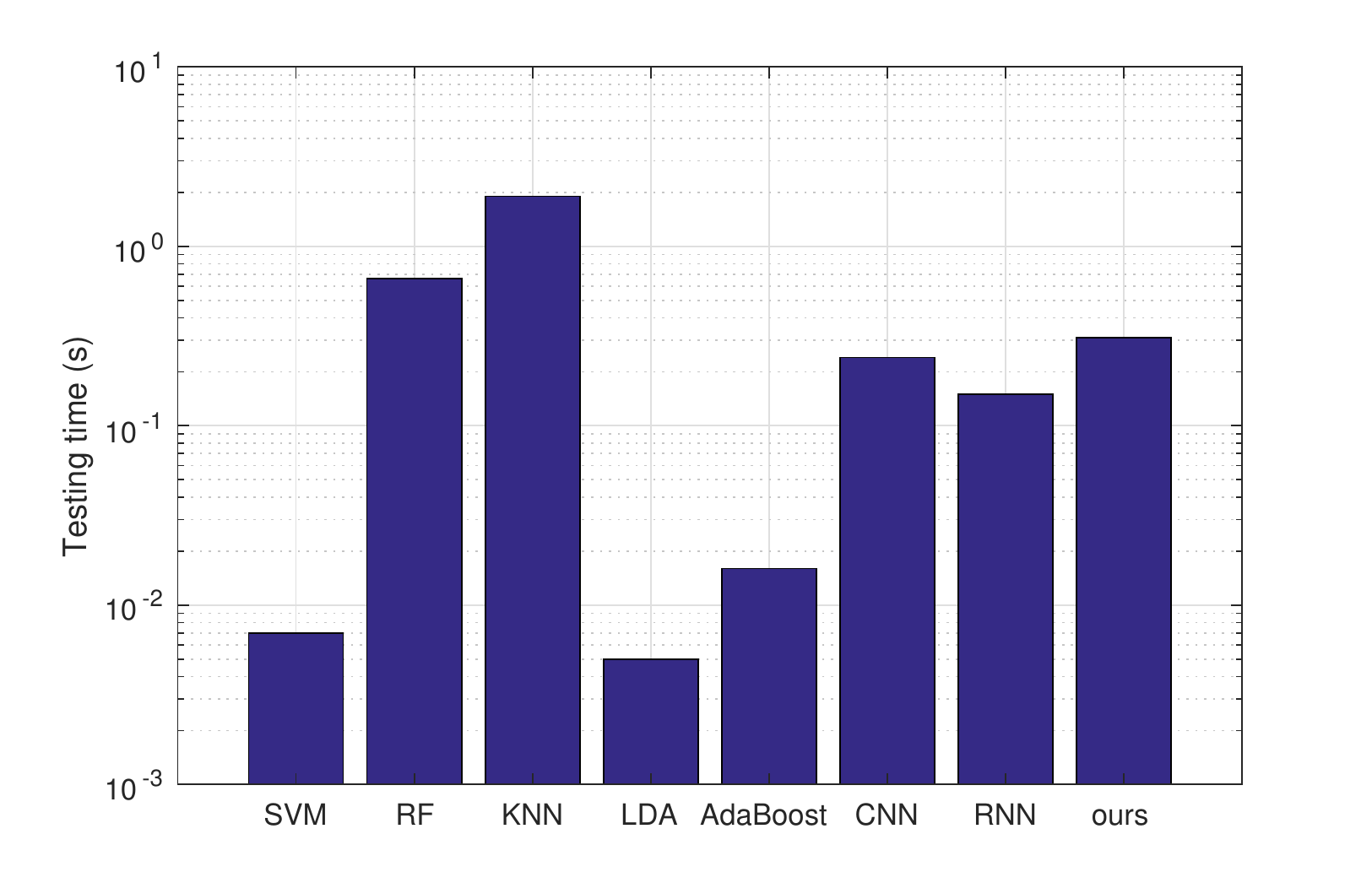} 
}
\caption{The accuracy and testing time comparison over \textit{emotiv} dataset}
\label{fig:acc_testtime}
\end{figure}

Subsequently, to demonstrate the efficiency of the proposed approach, we compare our method with the state-of-the-art methods and report the accuracy and testing time in Figure~\ref{fig:acc_testtime}. 

To conclude, our model still achieves good performance with EEG signals collected from hardware with fewer channels and in a more natural setting.

\begin{figure}[!]
\centering
\includegraphics[width=0.8\linewidth]{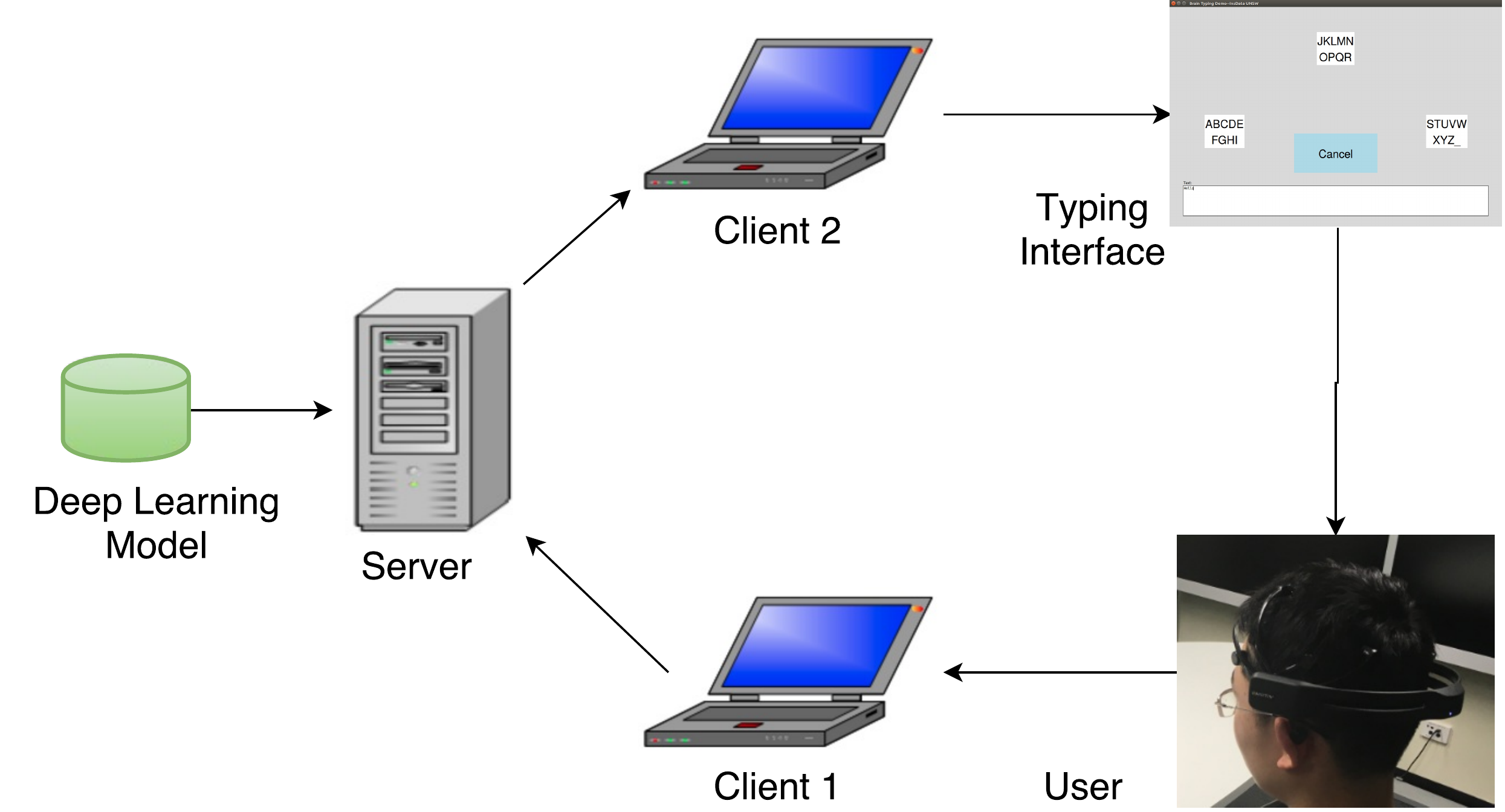}
\caption{Overview of the brain typing system. The user's typing intent is collected by headset and sent to the server through client 1. 
The server uses the pre-trained deep leaning model to recognize the intent, which is used to control the typing interface through client 2. The server and clients are connected using TCP connections.}
\label{fig:systemoverview}
\end{figure}

\section{APPLICATION: BRAIN TYPING SYSTEM} 
\label{sec:brain_typing_system}

Based on the high EEG signals classification accuracy of the proposed deep learning approach, in this section, we develop an online brain typing system to convert user's thoughts to texts. 

The brain typing system (as shown in Figure~\ref{fig:systemoverview}) consists of two components: the {\em pre-trained deep learning model} and the {\em online BCI system}. 
The pre-trained deep learning model, which is trained offline, aims to accurately and recognize the user's typing intent in real time.
 This model 
(introduced in detail in Section~\ref{sec:approach}) 
is central to the operation of the brain typing system, and is the main contribution of this paper.
The online system contains 5 components: the EEG headset, the client 1 
(data collector), the server, the client 2 (typing command receiver), and the typing interface. 

The user wears the Emotiv EPOC+ headset (introduced in Section~\ref{sub:case_study})
which collects EEG signals and sends
the data to 
client 1 through a bluetooth connection. The raw EEG signals are transported to the server through a TCP connection. The server 
feeds the incoming EEG signals to the pre-trained deep learning model. The model produces a 
classification 
decision and converts it to the corresponding typing command which is sent to 
client 2 
through a TCP connection. The typing interface receives the command 
and manifests the appropriate typing action.

\begin{figure}
\centering
\includegraphics[width=\linewidth]{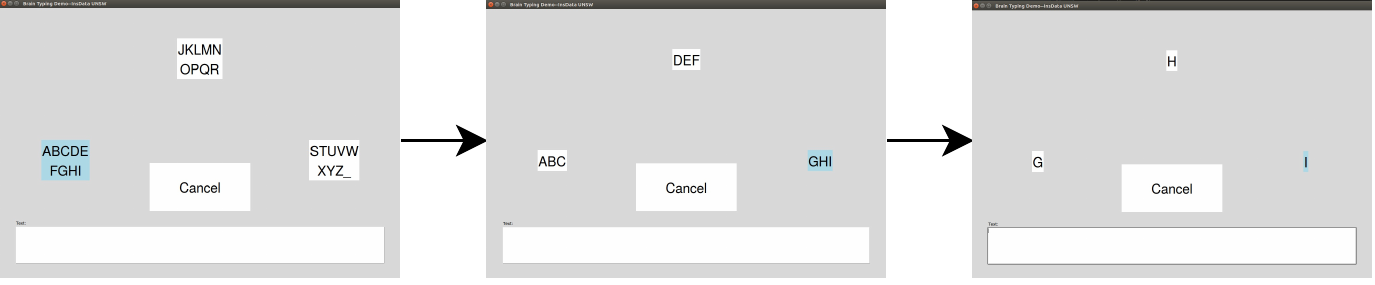}
\caption{The brain typing procedure to type the character `I'. Firstly, select the left character block (contains `ABCDEFGHI' characters) in the \textit{initial interface} and then confirm the selection to step in the corresponding \textit{sub-interface}; then, select the right character block (contains `GHI' characters) in the \textit{sub-interface} and confirm to jump to the \textit{bottom interface}; at last, select the right character block (only contains `I') and the character `I' will appear in the display block after the confirmation.}
\label{fig:typingprocedure}
\vspace{-5mm}
\end{figure}

Specifically, the typing interface (Figure~\ref{fig:typingprocedure}) can be divided into three levels: the initial interface, the sub-interface, and the bottom interface. 
All the interfaces have similar structure: three \textit{character blocks} (separately distributed in left, up, and down directions), a \textit{display block}, and a \textit{cancel button}.  
The display block shows the typed output and the cancel button is used to cancel the last operation. All 
interfaces include the display block and cancel button but differ in character blocks. 
The typing system totally includes $27=3*9$ characters (26 English alphabets and the 
space 
bar) and all of them are separated by 3 character blocks (each block contains 9 characters) in the initial interface. 
Overall, there are 3 alternative selections and each selection will lead to a specific sub-interface which contains 9 characters. 
Again, the $9=3*3$ characters are divided into 3 character blocks and each of them is connected to a bottom interface. In the bottom interface, each block represents only one character. As an example, Figure~\ref{fig:typingprocedure} shows the procedure to type the character `I'.

In the brain typing system, there are 5 commands to control the interface: `left', `up', `right', `cancel', and 
`confirm'.
 Each command corresponds to a specific motor imagery EEG category (as shown in Table~\ref{tab:table2}). To type each single character, the interface is supposed to accept 6 commands. 
Consider typing the letter `I' as an example (see Figure~\ref{fig:typingprocedure}). The sequence of commands to be entered is as follows:
`left', `confirm', `right', `confirm', `right', `confirm'. 
In our practical deployment, the sampling rate of Emotiv EPOC+ headset is set as 128Hz, which means the server can receive 128 EEG recordings each second. 
Since the brainwave signal varies rapidly and is very easy to be affected by noises, the EEG data stream is sent to server {\it each half second}, which means that the server receives 64 EEG samples each time. The 64 EEG samples are classified by the deep learning framework and generates 64 categories of intents. we calculate the mode of 64 intents and regard the mode as the final intent decision. 
Furthermore, 
to achieve steadiness and reliability, 
the server sends command to 
client 2 only if \textit{three consecutive decisions remain consistent}. After the command is sent, the command list will be reset and the system will wait until 3 consistent decisions are made.
Therefore, client 2 must wait for at least 1.5 seconds for a command and the entire process of typing each character takes at least 9 ($6*1.5$) seconds.
In other words, theoretically, the proposed brain typing system can achieve the highest typing speed of $6.67=60/9$ characters per minute. 

\section{DISCUSION} 
\label{sec:discusion}
The proposed deep learning framework achieves the highest accuracy compared to the state-of-the-art EEG classification methods. The classification accuracy of the public dataset (eegmmidb) is consistently higher than the local real-world dataset (emotiv). 
The possible reason may be due to the different channels of two datasets (eegmmidb contains 64 channels and emotiv only takes 14 channels). 
In general, our framework can achieve high classification accuracy with both datasets.

The accuracy in the online mode is however lower than what can be achieved in an offline setting (over 95\%), which could be attributed to a number of reasons.
At first, the user's mental state and fluctuations in emotions may affect the quality of the EEG signals. For example, if the offline dataset used to train the deep learning model is collected when the user is in an excited emotion state but then applied in an online setting when the user is upset, would lead to low accuracy.
In addition, subtle variations in the way the EEG headset is mounted on the subject's head may also impact online decision making. Specifically, the position of each of the electrodes (e.g.. the 14 electrodes in the Emotiv headset) on the scalp may vary during training and testing. Moreover, the EEG signals vary from person to person, which makes it difficult to construct a common model that applies to all individuals.
One of 
our future work is to identify the intra-class variabilities shared by all the activities of different subjects. Last but not least, some limitations are caused by the intrinsic attributes of the headset. For instance, the headset used in our case study is too tight for the user to wear longer than 30 minutes and the conductive quality of the wet electrodes decreases after prolonged usage.

\section{RELATED WORK}
\label{sec:related_work}

In EEG decoding and interpretation area, there are mainly two research directions: the {\em EEG feature representation} and the {\em EEG classifier}. 

Effectively representing features from EEG raw data
is 
critical for the classification accuracy for the complexity and high dimensionality of EEG signals.
Vézard et al. \cite{vezard2015eeg} employ common spatial pattern (CSP) along with LDA to pre-process the EEG data and obtain an accuracy of 71.59\% to binary alertness states. 
The autoregressive (AR) modeling approach, a widely used algorithm for EEG feature extraction, is also broadly combined with other feature extraction techniques to gain a better performance \cite{rahman2012comprehensive,zhang2016classification}. 
Duan et al. \cite{duan2016feature} introduce the Autoencoder method for feature extraction and finally obtain a classification accuracy of 86.69\%. 
Wavelet analysis \cite{albert2016automatic} 
is employed to carry on a diagnosis of Traumatic Brain Injury (TBI) by quantitative EEG (qEEG) data and reaches 87.85\% accuracy. 
Power spectral density \cite{al2017predicting} is extracted as EEG data features to input into SVM.
The work achieves 76\% accuracy with the data from FC4 $\sim$ AF8 channels and 92\% with the data from CPz $\sim$ CP2 channels.

Recently, more and more studies exploit deep learning \cite{jirayucharoensak2014eeg,ren2014convolutional} to classify EEG signals. 
The work in \cite{An2014} builds one deep belief net (DBN) classifier and achieves the accuracy of 83\% on binary classification.
The algorithm combined CNN and stacked Autoencoder (SAE) is investigated in \cite{tabar2016novel} to classify EEG Motor Imagery signals and results in 90\% accuracy. 

Based on the EEG signal decoding, a few researches start to explore the non-invasive brain typing method. The approach in \cite{speier2017online} enables ALS patients to type through BCI and achieves the typing rate 
of 6 characters per minute. The authors in \cite{moghadamfalahi2015language} investigate three kinds of typing interfaces and illustrates that both matrix presentation and RSVP (rapid serial visual presentation) can work well. 

\section{CONCLUSION AND FUTUREWORK}
\label{sec:conclusion}
In this paper, we present a hybrid deep learning model to decode the raw EEG signals for the aim of converting the user's thoughts to texts. 
The model employs the RNN and CNN to learn the temporal and spatial dependency features from the input EEG raw data and then stack them together. 
Our proposed approach adopts an 
Autoencoder to recognize the stacked feature and to eliminate the artifacts and employs the XGBoost classifier for the intent recognition. 
We evaluate our approach on a public MI-EEG dataset and also a real world dataset collected by ourselves. Both 
results (95.53\% and 94.27\%) outperform the state-of-the-art methods.

Our future work will focus on improving the accuracy in the {\em person-independent} scenario, wherein some subjects 
participate in the training 
and the rest of subjects 
involve in the testing. 
Our recent study on human activity recognition atop multi-task learning based framework \cite{yao2016learning} shows the capability to capture certain underlying local commonalities under the intra-class variabilities shared by all the activities of different subjects. 

\balance
\bibliographystyle{IEEEtran}
\bibliography{sigproc}

\end{document}